\documentclass[acmsmall]{acmart}
\AtBeginDocument{%
  }

\acmDOI{10.1145/3839470}

\usepackage{mathpartir}
\usepackage{mathtools}
\usepackage{xspace}
\usepackage{listings}
\usepackage{xcolor}  
\usepackage{cleveref}
\usepackage{stmaryrd}
\usepackage{subcaption}

\usepackage{ifthen}

\newboolean{pldiversion}
\setboolean{pldiversion}{false} 

\newtheorem{theorem}{Theorem}
\newtheorem{lemma}{Lemma}

\lstdefinestyle{cuda}{
  language=C,
  basicstyle=\ttfamily\footnotesize,  
  keywordstyle=\color{blue}\bfseries,
  commentstyle=\color{green!50!black},
  stringstyle=\color{orange},
  numbers=none,  
  tabsize=2,
  showstringspaces=false,
  breaklines=true,  
  breakatwhitespace=true,
  captionpos=b
}

\newcommand{\hookdownarrow}{%
  \mathrel{\rotatebox[origin=c]{-90}{$\hookrightarrow$}}}

\newcommand{\CC}{C\nolinebreak\hspace{-.05em}\raisebox{.4ex}{\small\bf +}\nolinebreak\raisebox{.4ex}{\small\bf +}}

\newcommand{\tool}{{\sc Volta}\xspace}
\newcommand{\lang}{\ensuremath{\mathcal{L}}}
\newcommand{\setidx}{\ensuremath{{I}}}

\newcommand{\kw}[1]{\textup{\textsf{\textbf{#1}}}} 
\newcommand{\tpred}[1]{\textit{#1}}
\newcommand{\cvar}{c} 
\newcommand{\rvar}{r} 
\newcommand{\eop}{\oplus} 
\newcommand{\svar}{s} 
\newcommand{\sseq}{\mathbin{\fatsemi}} 
\newcommand{\deref}{\mathord{*}}
\newcommand{\sassn}{\coloneqq} 
\newcommand{\sidx}[2]{#1[#2]} 
\newcommand{\tvar}{T} 
\newcommand{\pvar}{P} 
\newcommand{\pemp}{\diamond} 
\newcommand{\ppar}{\parallel} 

\begin{document}

\title{Equivalence Checking of ML GPU Kernels}

\author{Benjamin Driscoll}
\authornote{Both authors contributed equally to this research.}
\correspondingauthor
\orcid{0000-0002-1901-1880}
\affiliation{%
  \institution{Stanford University}
  \city{Stanford}
  \country{USA}
}
\email{bdrisc@cs.stanford.edu}

\author{Kshitij Dubey}
\authornotemark[1]
\orcid{0009-0000-1397-0661}
\affiliation{%
  \institution{Microsoft Research}
  \city{Bengaluru}
  \country{India}
}
\email{t-ksdubey@microsoft.com}

\author{Anjiang Wei}
\orcid{0000-0003-1654-6027}
\affiliation{%
  \institution{Stanford University}
  \city{Stanford}
  \country{USA}
}
\email{anjiang@cs.stanford.edu}

\author{Neeraj Kayal}
\orcid{0009-0002-0036-7702}
\affiliation{%
  \institution{Microsoft Research}
  \city{Bengaluru}
  \country{India}
}
\email{neeraka@microsoft.com}

\author{Rahul Sharma}
\authornote{This work was done while the author was at Microsoft Research.}
\orcid{0000-0001-7527-4653}
\affiliation{%
  \institution{Google DeepMind}
  \city{Bengaluru}
  \country{India}
}
\email{simplyrahul@google.com}

\author{Alex Aiken}
\orcid{0000-0003-3372-9036}
\affiliation{%
  \institution{Stanford University}
  \city{Stanford}
  \country{USA}
}
\email{aiken@cs.stanford.edu}


\begin{abstract}
With the rapid progress of deep learning and large language models (LLMs), companies spend enormous sums executing GPU kernels. These kernels have become prime targets for aggressive optimization. Recent efforts increasingly leverage LLMs to generate GPU kernels, but make no formal guarantees about the generated kernels. We present the first equivalence checker for GPU kernels and use it to formally verify the correctness of machine learning (ML) kernels optimized by hand, by LLM, and by compiler. We show that our equivalence checker is sound and, for a well-defined class of GPU kernels which includes many programs of interest, complete. Our implementation, \tool, can verify ML computations such as convolutions, matrix multiplications, and various attention mechanisms.
\end{abstract}




\maketitle

\section{Introduction}
\label{sec:intro}
Given two GPU kernels---a reference implementation and an optimized counterpart---our goal is to prove their extensional equivalence, i.e., that they produce identical outputs on all valid inputs, and, thereby, to verify the correctness of the optimized implementation.
Equivalence checking has been extensively studied for over half a century~\cite{mccarthy} and recent works, such as \citet{counter}, have successfully proven the correctness of optimized CPU code. However, GPU programming introduces unique challenges arising from fine-grained synchronization among a massive number of threads and, to date, the literature proffers no equivalence checker for GPU programs.

In the past few years, companies have begun to spend enormous sums executing the GPU kernels for deep learning and large language models. These kernels have become prime targets for aggressive optimization, performed fully automatically by agents~\cite{crfm,astra,kevin}, or even the old-fashioned way, manually~\cite{boehm} or by compilers~\cite{triton}. Recent efforts increasingly leverage large language models to generate GPU kernels. For example, NVIDIA has used DeepSeek-R1,\footnote{\url{https://developer.nvidia.com/blog/automating-gpu-kernel-generation-with-deepseek-r1-and-inference-time-scaling}} Stanford researchers have used OpenAI o3 and Gemini 2.5 Pro,\footnote{\url{https://crfm.stanford.edu/2025/05/28/fast-kernels.html}} the Kevin-32B model was trained with reinforcement learning for kernel generation,\footnote{\url{https://cognition.ai/blog/kevin-32b}} Feature Search and Reinforcement combines search with reinforcement learning to derive kernels~\cite{fsr}, Standard Kernel is building AI infrastructure for kernel generation,\footnote{\url{https://standardkernel.com}} and {\tt mako} offers an interface for generating GPU kernels with different LLMs.\footnote{\url{https://generate.mako.dev}} Some of these efforts, such as Standard Kernel and {\tt mako}, subject kernels to a range of empirical correctness checks, program analyses, and/or agent review, but none currently provides formal correctness guarantees. Such guarantees offer greater confidence than empirical validation alone, which can be valuable before deployment in a production setting, where bugs can be costly.

GPU kernels execute thousands of threads in parallel on runtimes such as NVIDIA's CUDA~\cite{cuda} or AMD's ROCm~\cite{rocm}, which can introduce subtle concurrency errors that testing alone cannot reliably detect. One major source of such errors is data races, which may appear only in extremely rare executions or may remain latent when kernels rely on implicit synchronization that happens only for a specific version of the test runtime or the test hardware but that is not guaranteed to hold in general. For example, a widely cited GPU programming tutorial~\cite{reduction} relies on warp-level synchronization that was removed after CUDA 9.0~\cite{implicitblog}. Testing-based correctness checking is further complicated by floating-point arithmetic. GPU optimizations often reorder arithmetic operations, causing small numerical differences between the outputs of the reference and optimized implementations. Such variations are usually tolerated in tests, yet similar deviations can arise from genuine semantic bugs. For instance, an unsound optimization that omits a clipping operation may seem acceptable for some test inputs even though it reduces overall model accuracy. While compilers rarely introduce such mistakes, they are plausible when kernels are hand-optimized by human programmers or generated by agents.

State-of-the-art equivalence checking supports single-threaded integer programs~\cite{counter} and high-level tensor operations~\cite{hec,tensorright}, but it is not obvious how to decompose the equivalence checking of GPU kernels into single-threaded sub-tasks. In an optimized kernel, there may be no one-to-one correspondence between its threads and the threads in the reference implementation. In addition, existing techniques have no support for synchronization among multiple threads.

Although equivalence checking is undecidable in general and GPUs allow programmers to write arbitrary code, efficient GPU kernels typically follow structural patterns that, incidentally, make formal reasoning feasible.
We assume that the programs analyzed by our equivalence checker are {\em structured-CTAs}. Importantly, the number of threads, the sizes of tensors, the targets of pointers, and the values of any expressions that are branched on must be statically known.
Although it is easy to construct programs that violate these assumptions and render our techniques inapplicable, we and others~\cite{test-amplification,weft,faialaa} observe that these assumptions very often hold in practice (Section~\ref{sec:practical}). For example, JAX, a popular ML framework, requires tensor dimensions to be fixed\footnote{\url{https://docs.jax.dev/en/latest/notebooks/Common_Gotchas_in_JAX.html\#dynamic-shapes}}; the underlying XLA compiler requires tensor sizes and data flow graphs to be known a priori to perform aggressive optimizations. We discuss structured-CTAs further in Section~\ref{sec:practical} and formally define them in Section~\ref{sec:abstraction}. Our implementation, \tool,\footnote{Verification Of Large Thread Arrays} checks that its input is a structured-CTA and raises an exception if it is not.
Finally, we note that, while our work focuses on kernels targeting NVIDIA GPUs due to their widespread adoption, the techniques are also applicable to other GPU runtimes.

The core of our equivalence checker is symbolic evaluation.
Consider a tensor computation---a workload that dominates machine learning. We symbolically execute the reference computation under an arbitrary, fixed schedule to derive an expression over $\mathbb{R}$ that represents the output tensor as a function of the input tensors, discovering any races or deadlocks as we go.
Our main technical result, {\em confluence}, which we formally verify in the \textsc{Agda} \cite{agda} proof assistant, guarantees that all symbolic evaluator schedules lead to the same expression or report a race or deadlock. The same procedure is applied to the optimized computation. Finally, we apply a procedure to decide equality of the two symbolic expressions and, if they are equal, we conclude that the optimized kernel is semantically equivalent to the reference kernel, for all schedules.

We are not aware of a proof in the literature that the symbolic identities we encounter are decidable, though there is work on related problems, e.g., \cite{realexp,li-wu-itcs26}. In particular, in Section~\ref{sec:decpro}, we prove decidability for equalities of the form $\sum_{i=1}^m p_i(\bar{x})e^{h_i(\bar{x})}=0$, given multivariate polynomials $p_1, h_1, \ldots \in \mathbb{R}[\bar{x}]$. 
It is worth noting that optimized kernels are almost never bit-wise equivalent to the reference implementation in floating-point arithmetic, since most optimizations reorder arithmetic operations when distributing work among GPU threads. 

Although there is no prior equivalence checker for GPU programs, there are prior verification frameworks  that check weaker properties such as data race freedom~\cite{faialaa}. Our equivalence checker, \tool, soundly verifies data race and deadlock freedom of kernels utilizing barrier synchronization (Theorem~\ref{thm:data-race-freedom}) in time proportional to the number of instructions. It satisfies a {\em true positives property}, which guarantees that every reported data race or deadlock corresponds to a concrete schedule that triggers erroneous behavior (Theorem~\ref{thm:true-positives-property}). Barrier synchronization operations include {\tt mma.sync} and {\tt syncwarp}, which programmers commonly employ when accelerating machine learning workloads with tensor cores~\cite{tc-wp}. We compare with data race checking tools in Section~\ref{sec:dr} and Section~\ref{sec:related}.

\tool checks the identities output by its symbolic executor in less than 3 minutes on all evaluated benchmarks, including convolution, matrix multiplication, and attention (Section~\ref{sec:evaluation}).
\tool operates as a black-box equivalence checker, requiring no information about how an optimized kernel was produced; it verifies correctness solely from the assembly corresponding to the optimized program itself. In particular, \tool analyzes PTX assembly, the lowest documented level of the stack for NVIDIA GPUs~\cite{ptx}. Operating at the PTX level enables our approach to handle various developer front-ends. For example, GPU programmers can write CUDA kernels or CUTLASS templates~\cite{cutlass}. NVIDIA's {\tt nvcc} compiler translates both into PTX.

In sum, we make the following contributions:
\begin{itemize}
    \item A linear time algorithm for detecting races and deadlocks for a class of GPU kernels, structured CTAs, utilizing barrier synchronization.
    \item A proof of decidability for equalities of the form $\sum_{i=1}^m p_i(\bar{x})e^{h_i(\bar{x})}=0$, given multivariate polynomials $p_1, h_1, \ldots \in \mathbb{R}[\bar{x}]$.
    \item The first equivalence checker for a practical class of GPU kernels (again, structured CTAs), \tool, that can verify the correctness of optimized convolutions, matrix multiplications, reductions, and attention mechanisms. We prove that \tool guarantees data race freedom and satisfies a true positives property, and we formalize the key, constituent lemma in \textsc{Agda}.
\end{itemize}

The rest of the paper is as follows. \Cref{sec:motivating} shows, at a high-level, how \tool verifies a motivating example. \Cref{sec:background} provides background on GPU architecture, synchronization, and ML GPU kernels. \Cref{sec:algo} describes our symbolic execution algorithm formally. \Cref{sec:formal} establishes our main technical results. \Cref{sec:impl} describes \tool's implementation and \Cref{sec:evaluation} evaluates it across a diverse set of GPU kernels, demonstrating its ability to verify realistic programs and detect bugs. \Cref{sec:related} reviews related work and \Cref{sec:conclusion} concludes with directions for future research.
\section{Motivating Example}
\label{sec:motivating}
\begin{figure}[h]
\centering
\begin{subfigure}[b]{0.45\linewidth}
\begin{lstlisting}[language=C, basicstyle=\ttfamily\scriptsize]
__global__ void
softmax (float *x, float *y, int N) {
    extern __shared__ float buf[];

    int tid = threadIdx.x;
    buf[tid] = expf(x[tid]);
    __syncthreads();

    float sum = 0.0f;
    for (int i = 0; i < N; i++)
        sum += buf[i];
    y[tid] = buf[tid] / sum;
    return;
}
\end{lstlisting}
\caption{Naive Softmax}
\label{lst:naive-softmax}
\end{subfigure}\hfill
\begin{subfigure}[b]{0.55\linewidth}
\begin{lstlisting}[language=C, basicstyle=\ttfamily\scriptsize]
__global__ void
online_softmax (float *x, float *y, int N) {
    int tid = threadIdx.x;

    float m = -INFINITY;  // running max
    float d = 0.0f;       // running denom

    for (int i = 0; i < N; i++) {
        float v = x[i];
        float new_m = fmaxf(m, v);
        d = d * expf(m - new_m) + expf(v - new_m);
        m = new_m;
    }
    y[tid] = expf(x[tid] - m) / d;
    return;
}
\end{lstlisting}
\caption{Online Softmax (FlashAttention)}
\label{lst:online-softmax}
\end{subfigure}
\caption{Naive softmax (left) requires materializing all exponentiations in shared memory; 
online softmax (right, used in FlashAttention) maintains a running max and normalization factor, 
enabling numerically stable and memory-efficient streaming.}
\label{fig:wex}
\end{figure}
As a motivating example, consider the $\mathit{softmax}$ computation, which appears in the attention mechanism of large language models (LLMs). Attention is the primary performance bottleneck during LLM inference with long contexts~\cite{fa}. Listing~\ref{lst:naive-softmax} presents a naive reference implementation of softmax in CUDA. In this example, we use $N=4$, meaning that four GPU threads operate on a vector $[x_1, x_2, x_3, x_4]$ to compute $[y_1, y_2, y_3, y_4]$, where
\begin{equation}
\label{eq:softmax}
y_i = e^{x_i} / \left({\textstyle \sum_{j=1}^{4} e^{x_j}}\right)
\end{equation}

The parameter $N$ is typically known at compile time. Kernels are often specialized for particular values of $N$, since a single CUDA kernel rarely achieves optimal performance across all possible configurations.

Listing~\ref{lst:naive-softmax} uses an array, {\tt buf}, stored in {\em shared memory}, a software-managed cache accessible to all threads within a thread block. Each thread has a unique identifier, {\tt tid}, and thread $i$ stores $e^{x_i}$ in {\tt buf}. The {\tt syncthreads} call is a barrier primitive that synchronizes all the threads, ensuring that all shared memory writes are visible before any thread proceeds.  After synchronization, each thread $i$ sums the elements of {\tt buf} into the variable {\tt sum} and writes $\frac{e^{x_i}}{\mathtt{sum}}$ to $y_i$.

Listing~\ref{lst:online-softmax} gives a softmax implementation used in FlashAttention~\cite{fa}, a major advance in the implementation of attention mechanisms. It employs an online softmax computation that is both numerically stable and avoids the use of shared memory. This formulation enables tiling optimizations that are key to FlashAttention's performance. The two softmax implementations shown in Figure~\ref{fig:wex} are not bit-wise identical due to floating-point rounding differences, yet both are intended to compute the mathematical expression in Equation~\ref{eq:softmax}. \tool is designed to verify that they both, in fact, compute this expression.

Conceptually, our analysis expands the {\tt softmax} kernel into a list of per-thread programs, where the $i^{th}$ program represents the code executed by thread $i$. 

\begin{figure}[h]
\centering
\scriptsize
\renewcommand{\arraystretch}{1.3}
\setlength{\tabcolsep}{6pt}
\begin{tabular}{c c c c}
\textbf{tid=0} & \textbf{tid=1} & \textbf{tid=2} & \textbf{tid=3} \\
buf[0] = exp(x[0]); & buf[1] = exp(x[1]); & buf[2] = exp(x[2]); & buf[3] = exp(x[3]); \\
\hline
\_\_syncthreads(); & \_\_syncthreads(); & \_\_syncthreads(); & \_\_syncthreads(); \\
\hline
sum = 0; & sum = 0; & sum = 0; & sum = 0; \\
sum += buf[0]; & sum += buf[0]; & sum += buf[0]; & sum += buf[0]; \\
sum += buf[1]; & sum += buf[1]; & sum += buf[1]; & sum += buf[1]; \\
sum += buf[2]; & sum += buf[2]; & sum += buf[2]; & sum += buf[2]; \\
sum += buf[3]; & sum += buf[3]; & sum += buf[3]; & sum += buf[3]; \\
y[0] = buf[0]/sum; & y[1] = buf[1]/sum; & y[2] = buf[2]/sum; & y[3] = buf[3]/sum; \\
\end{tabular}
\caption{Naive softmax execution for N=4 threads. Each column shows the code executed by a single thread indexed by {\bf tid}.}
\label{fig:naive-softmax-table}
\end{figure}

\Cref{fig:naive-softmax-table} illustrates the expanded program. Observe that the code executed by each thread is straight-line, containing no branches or loops. Moreover, the addresses corresponding to all memory accesses are statically known.

Note that this representation is shown only for clarity of exposition. In fact, the actual implementation (\Cref{sec:impl}) never constructs the fully expanded program, nor does it perform other AST transformations. It operates over the original program, requiring that values are sufficiently static such that the symbolic executor can resolve all control flow.

Given the program, \tool  performs simple round-robin scheduling: it iterates over each column until a thread encounters a {\tt syncthreads} barrier and blocks. For each thread, all shared memory locations that are read or written are recorded. When a thread is blocked, \tool proceeds to the next active thread. If the recorded accesses reveal any data races, \tool generates an error report. Once all threads reach the corresponding barrier, the recorded information is cleared, and all threads become active. If all threads become blocked, there is a deadlock and \tool also generates an error report. In Figure~\ref{fig:naive-softmax-table}, if the {\tt syncthreads} barrier is missing, \tool reports data races. For example, when analyzing the thread with {\bf tid} 0, the analysis records a read from {\tt buf[1]}. Subsequently, when analyzing the thread with {\bf tid} 1, it reports a data race between the read of {\tt buf[1]} in thread 0 and the write to {\tt buf[1]} in thread 1.

Given Listing~\ref{lst:naive-softmax}, the executor produces the following, symbolic output state:\footnote{We add the subscript ``naive'' in $y_\text{naive}$ to disambiguate it from the output of the optimized computation.}
\[
y_\text{naive}[0]=\frac{e^{x_0}}{D},\ y_\text{naive}[1]=\frac{e^{x_1}}{D},\ y_\text{naive}[2]=\frac{e^{x_2}}{D},\ y_\text{naive}[3]=\frac{e^{x_3}}{D},\ D=((e^{x_0} +e^{x_1})+e^{x_2})+e^{x_3}
\]

Confluence (Lemma~\ref{lem:confluence}) guarantees that, since we did not detect a data race, if we had used some other schedule different from the round-robin scheduling of \tool, we would still end up with an identical output state. Likewise, since we did not encounter a deadlock, no schedule will deadlock. Repeating the same process for the online softmax in Listing~\ref{lst:online-softmax} produces the following output:
\[
y_\text{opt}[0]=\frac{e^{x_0-M_3}}{D},\ y_\text{opt}[1]=\frac{e^{x_1-M_3}}{D},\ y_\text{opt}[2]=\frac{e^{x_2-M_3}}{D},\ y_\text{opt}[3]=\frac{e^{x_3-M_3}}{D}
\]
where
\[
M_1=\max(x_0,x_1),\ M_2=\max(M_1,x_2),\ M_3=\max(M_2,x_3)
\]
and
\[
D=\left(\left(e^{x_0-M_1}+e^{x_1-M_1}\right)e^{M_1-M_2}+e^{x_2-M_2}\right)e^{M_2-M_3}+e^{x_3-M_3}
\]
Finally, we create the following decision problem,
\[
\forall i\in\{0,1,2,3\}.\ y_\text{naive}[i]=y_\text{opt}[i]
\]

Our decision procedure (\Cref{sec:decpro}) successfully validates these identities, allowing \tool to conclude that the two programs in Figure~\ref{fig:wex} are equivalent. As an additional contribution, we demonstrate the decidability of such identities (Section~\ref{sec:decpro}).

\section{Background}
\label{sec:background}
We begin with necessary background on GPUs. To execute a task on a GPU, the CPU host launches a grid of cooperative thread arrays (CTAs), also known as {\em thread-blocks}. All CTAs share access to {\em global memory} (HBM), typically sized in gigabytes. Different CTAs in a grid usually write to disjoint regions of global memory due to the cost of inter-CTA synchronization, which was not even possible for many GPU generations. The GPU hardware schedules each CTA on a {\em streaming multiprocessor} (SM); a modern GPU may contain hundreds of SMs. 
The resources assigned to each SM are limited and hence each CTA is tasked to compute only a {\em tile} of the output tensor.
Each CTA consists of 128 to 1024 {\em threads}, depending on the hardware generation. Within a CTA, threads are grouped into {\em warps}, each consisting of 32 threads. Every thread maintains its own {\em registers} for local storage, while {\em shared memory} serves as a software-managed cache accessible to all threads in the same CTA.
To coordinate access to shared memory, threads must use explicit {\em synchronization} operations. Missing synchronization leads to data races and over synchronization leads to loss of performance. The shared memory available per CTA ranges from 48 KB to 256 KB, depending on the GPU architecture. In a typical CTA, threads first load data from global memory into shared memory, perform arithmetic operations while keeping intermediate values in registers or shared memory, and finally write the results back to global memory. 
Writing code for efficient CTAs is challenging because programmers must reason about massive thread-level parallelism and low-level hardware details. \tool aims to  statically identify incorrect or unsafe optimizations. We next describe the synchronization mechanisms available on GPUs.

\subsection{Synchronization Operations}
NVIDIA GPUs support independent thread scheduling, meaning that in the absence of explicit synchronization, each thread executes independently. The fundamental synchronization primitive is {\tt syncthreads}, a barrier that spans all threads within a CTA. Threads that reach the barrier earlier are suspended until all threads in the CTA arrive or exit, at which point execution resumes for all of the threads blocked at the barrier.

Another synchronization primitive is {\tt syncwarp}, which operates on a user-specified subset of threads within a single warp. This mechanism is commonly used when programming tensor cores---CUDA intrinsics such as {\tt wmma::mma\_sync} synchronize all threads in a warp to perform matrix multiplication. Most prior verification techniques for GPU kernels support only {\tt syncthreads} and therefore cannot reason about machine learning kernels that rely on warp-level synchronization. While {\tt syncthreads} is deadlock free (per the CUDA documentation, it can only be placed under conditionals that evaluate uniformly across the entire CTA), {\tt syncwarp} can introduce deadlocks.

In summary, the synchronization mechanisms we consider are barriers that block subsets of the available threads. Asynchronous producer-consumer-style primitives such as {\tt arrive} are beyond the scope of this paper. The race checking mechanism of \tool could potentially be swapped out for a more general, if less efficient, mechanism such as a happens-before analysis in the style of \citet{weft}, without disrupting the rest of the system, but we leave this for future work.

\subsection{Why Structured-CTAs?}
\label{sec:practical}
In this section, we discuss, at a high level, the assumptions that \tool makes about input kernels. We first state our assumptions, then motivate them. Note that these assumptions are checked: \tool raises an exception if a kernel does not satisfy our requirements (Section~\ref{sec:impl}).

To be checked for equivalence, a program must operate on a known number of CTAs and threads per CTA. Each CTA must operate over tensor tiles whose sizes are statically known. In each thread of each CTA, given the {\tt tid} (thread ID), all branch targets and addresses used in memory accesses must be statically known. We compare kernels CTA-to-CTA, so we do not support equivalence checking of optimizations that mix responsibilities between CTAs.
In Section~\ref{sec:abstraction}, we define a language of \textit{structured-CTAs}, formally encoding these assumptions, and, in Section~\ref{sec:formal}, prove the completeness of \tool for such programs.

Structured-CTAs reflect well-established best practices for achieving high GPU performance, as supported by empirical study. A large-scale analysis of GPGPU kernels in the wild found that most memory accesses are {\em control independent} and {\em data independent}~\cite{faialaa} (59.5\% of 2770 programs). Similarly, a study of the CUDA SDK showed that most programs exhibit {\em access invariance}~\cite{test-amplification}, meaning that threads perform the same memory accesses for all inputs (52 out of 76). {\sc Weft}~\cite{weft}, a race detector for scientific and cudaDMA kernels, makes all of the above assumptions---{\em all} 26 scientific computations studied by {\sc Weft} fall into the aforementioned class.

Moreover, we are focused on the kernels that are performance bottlenecks in the training or inference of machine learning (ML) models, such as matrix multiplications, convolutions, and multi-head attention. In these workloads, the structure of the ML model is fixed: tensor entries depend on runtime inputs, but tensor sizes are statically known. Kernels are therefore specialized for particular tensor dimensions to achieve maximum performance. In fact, JAX~\cite{googlejax}, a popular ML framework, only permits tensors with fixed dimensions.

Once tensor shapes are known, loop bounds can also be determined, since the bounds in tensor operations depend only on tensor shapes. \tool can statically evaluate these loops to ``flat'' expressions that do not contain conditionals, effectively unrolling them. 
We also require that the targets of conditional branches can be statically resolved; note this restriction  exploits the fact that efficient machine learning kernels do not make truly data dependent runtime decisions.  Under these assumptions, 
the code executed by each individual thread can be represented as {\em straight-line}, no dynamic branches or loops. It is common for threads to branch on predicates of their thread index or of a loop index (counting up to some tensor dimension); however, such control flow still preserves the straight-line property for each thread's execution. For example, causal masking in FlashAttention~\cite{fa} branches on whether an array index is above or below the diagonal, but this is a static condition given the tensor shape.

Efficient GPU kernels carefully orchestrate memory transfers by coalescing accesses to global memory and avoiding shared memory bank conflicts. As a result, GPU programmers typically define memory access patterns statically, making memory addresses independent of the tensor contents. Consequently, when a straight-line thread accesses an array element, both the base array and the exact index can be determined statically. This property allows the symbolic evaluator to record memory accesses precisely and to detect data races accurately.

Typically, ML kernels launch a grid of CTAs where the code of each CTA is identical and differs only in the instantiation of free variables {\tt blockIdx.x}, {\tt blockIdx.y}, and {\tt blockIdx.z}. Each CTA is responsible for computing a fixed-size {\em tile} of the output tensor. Many GPU optimizations focus on this CTA-level code. For instance, the CTA code can be modified to use tensor cores. When checking a reference kernel against an optimized kernel, we focus on proving equivalence between the code of a CTA from the reference kernel and the code of the corresponding CTA in the optimized kernel. Hence, correctness checking of optimizations that change the number of CTAs launched in a grid is beyond the scope of this work.
Since each pair of CTAs can be checked independently of one another, following prior work~\cite{weft}, our evaluation simply checks the equivalence of the CTAs with {\tt blockIdx.x} = {\tt blockIdx.y} = {\tt blockIdx.z} = 0 in the reference and the optimized kernels.

Which practical kernels fall outside the class of structured-CTAs? Notably, top-$k$/argsort kernels do, as commonly used for expert selection in mixture-of-experts models, beam search, and ranking. Their control flow is data dependent---comparison results determine swap operations---and implementations often use atomic operations for parallel reductions across CTAs.


Finally, a sound equivalence checker operating over floating-point arithmetic would likely reject many aggressively optimized kernels; IEEE-754 is not even associative. In practice, both developers and domain-specific compilers treat tensor values as real numbers when applying optimizations. Such transformations, similar in spirit to the {\tt --ffast-math} option in {\tt gcc}, are unsound for floating-point arithmetic. \tool follows the same convention as existing optimizers and models tensor elements as reals.



\section{Algorithm}
\label{sec:algo}

Our equivalence checker proceeds in two phases. The first phase symbolically executes a pair of programs, determining whether they are data race free and, if so, expresses their output values as terms of symbolic input tensors. If both programs are data race free, we proceed to the second phase, which verifies that, for each output element, the symbolic terms derived for both programs are equal. For example, when analyzing Listing~\ref{lst:naive-softmax}, the symbolic evaluator starts with environment \(\{\sidx{x}{0} \sassn \alpha, \sidx{x}{1} \sassn \beta, \sidx{x}{2} \sassn \gamma, \sidx{x}{3} \sassn \delta\}\) for fresh symbols $\alpha, \beta, \gamma, \delta$. It determines there is no data race and outputs a symbolic state with $\sidx{y}{0} \sassn e^\alpha / (e^\alpha+e^\beta+e^\gamma+e^\delta)$.

\begin{figure}[t]
\begin{alignat*}{4}
    &\text{Statement}\quad &&\svar\quad &&\Coloneqq\quad && \rvar \sassn \cvar
    \mid \rvar \sassn \rvar_1 \eop \rvar_2
    \mid \rvar \sassn \rvar'
    \mid \rvar \sassn \deref g
    \mid \deref g \sassn \rvar
    \mid \kw{sync}\, \setidx \\
    &\text{Thread Program}\quad &&\tvar &&\Coloneqq\quad &&\kw{return}
    \mid \svar \sseq \tvar \\
    &\text{Program}\quad &&\pvar &&\Coloneqq\quad &&\pemp
    \mid \pvar \ppar \tvar
\end{alignat*}
    \caption{Syntax of \lang{}, the language of structured-CTAs.}
    \label{fig:syntax}
\end{figure}

\subsection{Formal Setting}
\label{sec:abstraction}

As a setting to formalize our analysis, we develop a simple language \lang{} of structured-CTAs capturing the relevant aspects of PTX. The syntax of \lang{} programs is given in Figure~\ref{fig:syntax} in terms of registers $\rvar$, constants $\cvar$, shared memory addresses $g$, and sets of thread IDs (TIDs) $\setidx$. We give \lang{} a dynamics via a pair of small-step relations, $(\mathcal{G}, \mathcal{R}, P) \to (\mathcal{G}, \mathcal{R}, P)$ and $(\mathcal{G}, R, T) \to (\mathcal{G}, R, T)$:

\begin{mathpar}
    \inferrule[Schd]{
        (\mathcal{G}, \mathcal{R}(i), P(i)) \to (\mathcal{G}', R', T')
    }{
        (\mathcal{G}, \mathcal{R}, P) \to (\mathcal{G}', \mathcal{R}[i \mapsto R'], P[i \mapsto T'])
    }
\end{mathpar}
\begin{mathpar}
    \inferrule[Sync]{
        \forall i \notin I.\ P'(i) = P(i) \\
        \forall i \in I.\ (P(i) = \kw{return} \land P'(i) = \kw{return}) \lor (\exists T_i.\ P(i) = \kw{sync}\, I \sseq T_i \land P'(i) = T_i)
    }{
        (\mathcal{G}, \mathcal{R}, P) \to (\mathcal{G}, \mathcal{R}, P')
    }
\end{mathpar}
\begin{mathpar}
    \inferrule[Const]{
    }{
        (\mathcal{G}, R, r \coloneqq c \sseq T) \to (\mathcal{G}, R[r \mapsto c], T)
    } \and
    \inferrule[BinOp]{
    }{
        (\mathcal{G}, R, r \coloneqq r_1 \eop r_2 \sseq T) \to (\mathcal{G}, R[r \mapsto R(r_1) \eop R(r_2)], T)
    } \and
    \inferrule[RdReg]{
    }{
        (\mathcal{G}, R, r \coloneqq r' \sseq T) \to (\mathcal{G}, R[r \mapsto R(r')], T)
    } \and
    \inferrule[RdMem]{
    }{
        (\mathcal{G}, R, r \coloneqq \deref g \sseq T) \to (\mathcal{G}, R[r \mapsto \mathcal{G}(g)], T)
    } \and
    \inferrule[WrMem]{
    }{
        (\mathcal{G}, R, \deref g \coloneqq r \sseq T) \to (\mathcal{G}[g \mapsto R(r)], R, T)
    }
\end{mathpar}

The top-level small-step relation is indexed by configurations $(\mathcal{G}, \mathcal{R}, P)$ where $\mathcal{G}$ models shared memory and is a map from addresses to values, $\mathcal{R}$ models the threads' register files and is a map from TIDs to registers to values, and $P$ is the current program. We will consider both concrete program executions, where values are drawn from $\mathbb{R}$, and abstract program executions, where values are symbolic.

\textsc{Schd} allows any thread $i$ (which is not blocked at a \kw{sync}) to be advanced, modeling the non-deterministic execution of the machine. We use $f[x \mapsto y]$ to denote the function which takes value $y$ at $x$ and is $f$ elsewhere. The second top-level rule, \textsc{Sync}, advances the threads blocked at a $\kw{sync}\, I$, but only if every thread in $I$ is at a $\kw{sync}\, I$ or \kw{return}. When $\setidx = \{1, \ldots, N\}$ for $N$ the size of a thread block, \kw{sync} models the PTX instruction {\tt syncthreads}. When $\setidx \subseteq \{k, \ldots, k + 31\}$ with $k \bmod 32 = 1$, \kw{sync} models the warp-level barriers, e.g., {\tt syncwarp} with a mask or tensor core {\tt mma.sync}. Permitting threads in $I$ to be at a \kw{return} captures the flexibility of PTX, which allows threads to terminate early. The program is stuck if a thread $i_1$ synchronizes on $I_1$ and another $i_2$ on $I_2$, where $I_1 \ne I_2$ and $i_1, i_2 \in I_1 \cap I_2$. Although some GPU architectures may permit specific combinations of $I_1$ and $I_2$ to proceed without deadlock, we conservatively treat all such cases as stuck. Finally, note that \kw{sync} operations are not named: as with {\tt syncthreads} in CUDA, whichever \kw{sync} instances happen to align in the dynamics are matched with one another. (In fact, because control flow in a structured-CTA cannot depend on runtime values, the same instances align in every execution.)

The other small-step relation given above is a thread-level relation indexed by configurations $(\mathcal{G}, R, T)$. $\mathcal{G}$ is, again, a map from addresses to values, modeling shared memory, $R$ is the current thread's register file, and $T$ is the current thread's program. \textsc{Const}, \textsc{BinOp}, \textsc{RdReg}, \textsc{RdMem}, and \textsc{WrMem} perform the obvious updates to $\mathcal{G}$ and $R$. $\eop$ models the various arithmetic operations handled by the real implementation, \tool. Note that there is no thread-level small-step rule for a thread program of the form $\kw{sync}\, I \sseq T'$---one must use the top-level \textsc{Sync} rule.

Note that, for confluence (Lemma~\ref{lem:confluence}), it will be essential that (dynamic) control flow cannot affect whether a \kw{sync} executes or the set of addresses read and written, which is why thread programs are straight-line sequences of statements. That said, the formalism is more permissive than it may first appear: the operators $\eop$ are black boxes, assumed only to be pure functions of their register operands, and can therefore ``hide'' control flow. For example, $\eop$ could be $\max$ implemented via a branch.

\subsection{Symbolic Execution}
\label{sec:symex}
A \textit{program trace} is a derivation of the reflexive, transitive closure of $(\mathcal{G}, \mathcal{R}, P) \to (\mathcal{G}, \mathcal{R}, P)$, which we denote $(\mathcal{G}, \mathcal{R}, P) \to^* (\mathcal{G}, \mathcal{R}, P)$. A \textit{data race} occurs between two memory accesses $a_i$ and $a_j$ when at least one is a write (a \textsc{Schd} of a \textsc{WrMem}) and there exist two traces, one in which $a_i$ is directly followed by $a_j$ and another in which $a_j$ is directly followed by $a_i$. This is the definition used by, e.g., \citet{BROOKES2007227}.\footnote{We could define a data race without requiring that $a_i$ and $a_j$ follow each other directly. There exist traces in which they occur in different orders and follow each other directly iff there exist traces in which they occur in different orders.} Since data races are typically bugs and it is not clear, in general, how equivalence should be defined for programs that race, we extend the dynamics of \lang{} to return error value $\bot$ if the program has a data race and use this extended dynamics when performing symbolic execution.

Concretely, we maintain an extra context $\mathcal{X}$ tracking: (1) for each address $g$ and each thread $i$, all threads $j$ that have not performed a \kw{sync} with $i$ since it last read $g$; (2) for each address $g$, the last writer $i$ and all threads $j$ that have not performed a \kw{sync} with $i$ since it wrote $g$. That is,
\begin{align*}
    &\mathcal{X} : \text{MemEvs} \triangleq \text{Addr} \to (\text{TID} \to \mathscr{P}(\text{TID})) \times (\text{TID} \times \mathscr{P}(\text{TID}))
\end{align*}
Races arise because a thread reads an address before synchronizing with its last writer, or writes an address before synchronizing with its previous readers and its last writer. We define a predicate $\tpred{noRacingRd}$ taking a thread $i$ and the read data of $\mathcal{X}(g)$ and determining whether $i$ has synchronized with all other readers of $g$ as required. Likewise, we define a predicate $\tpred{noRacingWr}$ taking a thread $i$ and the write data of $\mathcal{X}(g)$ and determining whether $i$ is or has synchronized with the last writer of $g$ as required.
\begin{align*}
    &\tpred{noRacingRd} : \text{TID} \to (\text{TID} \to \mathscr{P}(\text{TID})) \to \text{Prop} \\ 
    &\tpred{noRacingRd}(i, \mathrm{rd}) \triangleq \forall j.\ i = j \lor i \notin \mathrm{rd}(j) \\
    &\tpred{noRacingWr} : \text{TID} \to \text{TID} \times \mathscr{P}(\text{TID}) \to \text{Prop} \\
    &\tpred{noRacingWr}(i, (j, J)) \triangleq i = j \lor i \notin J
\end{align*}
{\emergencystretch=2em We can now refine our thread-level small-step relation. The new relation $(\mathcal{X}, \mathcal{G}, R, T) \hookrightarrow_i (\mathcal{X}', \mathcal{G}', R, T)$ tracks the TID of the thread $i$ being executed, and includes an $\mathcal{X}$ in its configurations. The non-trivial changes are to \textsc{RdMem} and \textsc{WrMem}. Letting $\mathbb{I}$ be the set of all TIDs:\par}
\begin{mathpar}
    \inferrule[RdMem']{
        (\mathrm{rd}, \mathrm{wr}) = \mathcal{X}(g) \\
        \tpred{noRacingWr}(i, \mathrm{wr}) \\
        x' = (\mathrm{rd}[i \mapsto \mathbb{I}], \mathrm{wr})
    }{
        (\mathcal{X}, \mathcal{G}, R, r \coloneqq \deref g \sseq T) \hookrightarrow_i (\mathcal{X}[g \mapsto x'], \mathcal{G}, R[r \mapsto \mathcal{G}(g)], T)
    } \and
    \inferrule[WrMem']{
        (\mathrm{rd}, \mathrm{wr}) = \mathcal{X}(g) \\
        \tpred{noRacingRd}(i, \mathrm{rd}) \\
        \tpred{noRacingWr}(i, \mathrm{wr}) \\
        x' = (\mathrm{rd}, (i, \mathbb{I}))
    }{
        (\mathcal{X}, \mathcal{G}, R, \deref g \coloneqq r \sseq T) \hookrightarrow_i (\mathcal{X}[g \mapsto x'], \mathcal{G}[g \mapsto R(r)], R, T)
    }
\end{mathpar}
Next, we must adjust \textsc{Sync} so that $\mathcal{X}$ is updated appropriately:
\begin{gather*}
    \inferrule[Sync']{
        \forall i \notin I.\ P'(i) = P(i) \\
        \forall i \in I.\ (P(i) = \kw{return} \land P'(i) = \kw{return}) \lor (\exists T_i.\ P(i) = \kw{sync}\, I \sseq T_i \land P'(i) = T_i)
    }{
        (\mathcal{X}, \mathcal{G}, \mathcal{R}, P) \hookrightarrow (\tpred{syncMem}(I, \mathcal{X}), \mathcal{G}, \mathcal{R}, P')
    } \\ \\
\begin{alignedat}{2}
    &\tpred{syncMemRd}(I, \mathrm{rd})(i) &&\triangleq (\mathrm{rd}(i) \setminus I) \text{ if } i \in I \text{ else } \mathrm{rd}(i) \\
    &\tpred{syncMemWr}(I, (j, J)) &&\triangleq (j, J \setminus I) \text{ if } j \in I \text{ else } (j, J) \\
    &\tpred{syncMem}(I, \mathcal{X})(g) &&\triangleq (\tpred{syncMemRd}(I, \pi_1(\mathcal{X}(g))), \tpred{syncMemWr}(I, \pi_2(\mathcal{X}(g))))
\end{alignedat}
\end{gather*}
Finally, we would like symbolic execution to error when a race is detected rather than getting stuck:
\begin{mathpar}
    \inferrule[RdMemBad]{
        (\mathrm{rd}, \mathrm{wr}) = \mathcal{X}(g) \\
        \lnot \tpred{noRacingWr}(i, \mathrm{wr})
    }{
        (\mathcal{X}, \mathcal{G}, R, r \coloneqq \deref g \sseq T) \hookrightarrow_i \bot
    } \and
    \inferrule[WrMemBad]{
        (\mathrm{rd}, \mathrm{wr}) = \mathcal{X}(g) \\
        \lnot \tpred{noRacingRd}(i, \mathrm{rd}) \lor \lnot \tpred{noRacingWr}(i, \mathrm{wr})
    }{
        (\mathcal{X}, \mathcal{G}, R, \deref g \coloneqq r \sseq T) \hookrightarrow_i \bot
    } \and
    \inferrule[SchdBad]{
        (\mathcal{X}, \mathcal{G}, \mathcal{R}(i), P(i)) \hookrightarrow_i \bot
    }{
        (\mathcal{X}, \mathcal{G}, \mathcal{R}, P) \hookrightarrow \bot
    }
\end{mathpar}

We write the reflexive, transitive closure of $(\mathcal{X}, \mathcal{G}, \mathcal{R}, P) \hookrightarrow (\mathcal{X}', \mathcal{G}', \mathcal{R}', P')$ as $(\mathcal{X}, \mathcal{G}, \mathcal{R}, P) \hookrightarrow^* (\mathcal{X}', \mathcal{G}', \mathcal{R}', P')$.

Note that the formalism assumes all addresses are valid, though validity could be tracked by enriching the codomain of $\mathcal{G}$ and $R$ and adding a check in \textsc{RdMem'} and \textsc{WrMem'}.

\subsection{Guarantees}
\label{sec:formal}
We prove the soundness (Theorem~\ref{thm:sound-equiv-checking}) and completeness (Theorem~\ref{thm:complete-equiv-checking}) of our symbolic evaluation based equivalence checking. 
For soundness, we prove that if execution of two programs under the \textit{checked} dynamics $\cdot \hookrightarrow \cdot$ with symbolic inputs produces equal terms, then those programs evaluate to the same expression under the usual, \textit{unchecked} dynamics $\cdot \to \cdot$ for all inputs in $\mathbb{R}$. The central challenge lies in reasoning about the apparent non-determinism introduced by \textsc{Schd'}.
Since symbolic execution selects an arbitrary unblocked thread of the program for execution at each step, we must show that, for all programs, all possible schedules either produce the same output $\mathcal{G}, \mathcal{R}$, raise an error $\bot$ (indicating a data race), or become stuck (indicating a deadlock). Our key lemma is the following, which we verify in the \textsc{Agda} \citep{agda} proof assistant:
\begin{lemma}[Confluence]
\label{lem:confluence}
If $(\mathcal{X}, \mathcal{G}, \mathcal{R}, P) \hookrightarrow^* (\mathcal{X}_1, \mathcal{G}_1, \mathcal{R}_1, P_1)$ and $(\mathcal{X}, \mathcal{G}, \mathcal{R}, P) \hookrightarrow^* (\mathcal{X}_2, \mathcal{G}_2, \mathcal{R}_2, P_2)$ then there exists $\mathcal{X}', \mathcal{G}', \mathcal{R}', P'$ such that $(\mathcal{X}_1, \mathcal{G}_1, \mathcal{R}_1, P_1) \hookrightarrow^* (\mathcal{X}', \mathcal{G}', \mathcal{R}', P')$ and $(\mathcal{X}_2, \mathcal{G}_2, \mathcal{R}_2, P_2) \hookrightarrow^* (\mathcal{X}', \mathcal{G}', \mathcal{R}', P')$.
\end{lemma}
\begin{proof}
In our \textsc{Agda} development, we apply the standard technique for obtaining confluence results. Namely, we show a one-step confluence/diamond property and use it to tile the desired, many-step diamond. While $\cdot \hookrightarrow \cdot$ does not satisfy a one-step diamond property (consider the case where one of the initial steps is \textsc{SchdBad}), its reflexive closure does, which suffices. The proof has a large number of cases, but most boil down to commutativity lemmas such as: \textsc{Sync'} for $I$ followed by \textsc{Sync'} for $J$ produces the same configuration as \textsc{Sync'} for $J$ followed by \textsc{Sync'} for $I$, assuming all necessary hypotheses for these \textsc{Sync'} steps hold.
\end{proof}

We will use confluence to prove soundness and completeness of equivalence checking. But first, we take a brief detour to prove data race freedom and the true positives property for $\cdot \hookrightarrow \cdot$, our checked dynamics.

We will need some notation. Let $(\mathcal{G}, \mathcal{R}, P) \downarrow (\mathcal{G}', \mathcal{R}')$ denote that $(\mathcal{G}, \mathcal{R}, P) \to^* (\mathcal{G}', \mathcal{R}', \kw{return} \ppar \cdots \ppar \kw{return})$, $(\mathcal{G}, \mathcal{R}, \mathcal{X}, P) \hookdownarrow (\mathcal{G}', \mathcal{R}')$ denote that $(\mathcal{G}, \mathcal{R}, \mathcal{X}, P) \hookrightarrow^* (\mathcal{X}', \mathcal{G}', \mathcal{R}', \kw{return} \ppar \cdots \ppar \kw{return})$ for some $\mathcal{X}'$, and $(\mathcal{G}, \mathcal{R}, \mathcal{X}, P) \hookdownarrow \bot$ denote that $(\mathcal{G}, \mathcal{R}, \mathcal{X}, P) \hookrightarrow^* \bot$. Let $\mathcal{X}_\varnothing$ denote the initial context of memory events $\lambda\, \_.\, ((\lambda\, \_.\, \varnothing), (0, \varnothing))$ (the choice of $0$ here is arbitrary), and let $(\mathbb{V}, \eop_\mathbb{V})$ denote the collection (free algebra) of symbolic terms. We will sometimes drop the subscript on $\eop_\mathbb{V}$ where it is obvious. Finally, we define a \textit{map} $\varphi : (A, \eop_A) \to (B, \eop_B)$ to be a function $\varphi : A \to B$ between the collections of program values $A$ and $B$ such that \(\varphi(v_1 \eop_A v_2) = \varphi(v_1) \eop_B \varphi(v_2)\).

Now we are ready to proceed with our proofs. For every unchecked trace, there is a checked trace that reaches the same result, or detects a data race:
\begin{lemma}
\label{lem:dde154e4-0a35-4e5f-998a-804455feda48}
    If $(\mathcal{G}, \mathcal{R}, P) \to^* (\mathcal{G}', \mathcal{R'}, P')$ then, for all $\mathcal{X}$, $(\mathcal{X}, \mathcal{G}, \mathcal{R}, P) \hookrightarrow^* (\mathcal{X}', \mathcal{G}', \mathcal{R}', P')$  for some $\mathcal{X}'$ or $(\mathcal{X}, \mathcal{G}, \mathcal{R}, P) \hookrightarrow^* \bot$
\end{lemma}
\begin{proof}
    By induction on $(\mathcal{G}, \mathcal{R}, P) \to^* (\mathcal{G}', \mathcal{R}', P')$.
\end{proof}

Conversely, for every checked trace, there is an unchecked trace that reaches the same result.
\begin{lemma}
\label{lem:678b930f-c61b-4fe8-b686-eaec81d650d5}
    If $(\mathcal{X}, \mathcal{G}, \mathcal{R}, P) \hookrightarrow^* (\mathcal{X}', \mathcal{G}', \mathcal{R}', P')$ then $(\mathcal{G}, \mathcal{R}, P) \to^* (\mathcal{G}', \mathcal{R}', P')$.
\end{lemma}
\begin{proof}
    By induction on $(\mathcal{X}, \mathcal{G}, \mathcal{R}, P) \hookrightarrow^* (\mathcal{X}', \mathcal{G}', \mathcal{R}', P')$.
\end{proof}

The initial values of shared memory and registers do not affect whether a data race is detected:
\begin{lemma}
\label{lem:1b7311f3-3b17-43e3-ba53-4c81684cd977}
    If $(\mathcal{X}, \mathcal{G}_1, \mathcal{R}_1, P) \hookdownarrow \bot$ then $(\mathcal{X}, \mathcal{G}_2, \mathcal{R}_2, P) \hookdownarrow \bot$.
\end{lemma}
\begin{proof}
    By induction we can prove that, for all $\mathcal{X}, \mathcal{G}_1, \mathcal{R}_1, P, \mathcal{X}', \mathcal{G}_1', \mathcal{R}_1', P', \mathcal{G}_2, \mathcal{R}_2$, if we have $(\mathcal{X}, \mathcal{G}_1, \mathcal{R}_1, P) \hookrightarrow^* (\mathcal{X}', \mathcal{G}_1', \mathcal{R}_1', P')$, then $(\mathcal{X}, \mathcal{G}_2, \mathcal{R}_2, P) \hookrightarrow^* (\mathcal{X}', \mathcal{G}_2', \mathcal{R}_2', P')$ for some $\mathcal{G}_2', \mathcal{R}_2'$. Hence, the desired result follows from the fact that \textsc{RdMemBad} and \textsc{WrMemBad} depend only on the context of memory events and the program.
\end{proof}

The next theorem proves that if there is a data race, then the symbolic evaluation detects it:
\begin{theorem}[Data Race Freedom]
\label{thm:data-race-freedom}
     If $P$ has a data race, then $(\mathcal{X}, \mathcal{G}, \mathcal{R}, P) \hookdownarrow \bot$.
\end{theorem}
\begin{proof}
    The fact that $P$ has a data race implies there are two traces---that is, derivations of $(\mathcal{G}_o, \mathcal{R}_o, P) \to^* (\mathcal{G}', \mathcal{R}', P')$ for some $\mathcal{G}_o$, $\mathcal{R}_o$, $\mathcal{G}'$, $\mathcal{R}'$, $P'$---with a common prefix $\alpha$ followed by conflicting accesses $a_i, a_j$ in one case, and, in the other, $a_j, a_i$. By Lemma~\ref{lem:dde154e4-0a35-4e5f-998a-804455feda48} on $\alpha$, either we can obtain a derivation of $(\mathcal{X}, \mathcal{G}_o, \mathcal{R}_o, P) \hookrightarrow^* (\mathcal{X}_\alpha, \mathcal{G}_\alpha, \mathcal{R}_\alpha, P_\alpha)$ for some $\mathcal{X}_\alpha$, $\mathcal{G}_\alpha$, $\mathcal{R}_\alpha$, $P_\alpha$, or $(\mathcal{X}, \mathcal{G}_o, \mathcal{R}_o, P) \hookdownarrow \bot$ and, by Lemma~\ref{lem:1b7311f3-3b17-43e3-ba53-4c81684cd977}, we are done. In the former case, without loss of generality, let $a_j$ be a write. Either we can append $a_i$ to our $(\mathcal{X}, \mathcal{G}_o, \mathcal{R}_o, P) \hookrightarrow^* (\mathcal{X}_\alpha, \mathcal{G}_\alpha, \mathcal{R}_\alpha, P_\alpha)$ without obtaining $\bot$, or we are done by Lemma~\ref{lem:1b7311f3-3b17-43e3-ba53-4c81684cd977}. In the former case, the write $a_j$ now induces a $\lnot \tpred{noRacingRd}$ if $a_i$ is a read or a $\lnot \tpred{noRacingWr}$ if $a_i$ is a write, completing the proof by Lemma~\ref{lem:1b7311f3-3b17-43e3-ba53-4c81684cd977}.
\end{proof}

The next theorem proves that if the symbolic evaluation detects a data race, then there is actually a data race:
\begin{theorem}[True Positives Property]
\label{thm:true-positives-property}
     If $(\mathcal{X}_\varnothing, \mathcal{G}, \mathcal{R}, P) \hookdownarrow \bot$ then $P$ has a data race.
\end{theorem}
\begin{proof}
    Since $(\mathcal{X}_\varnothing, \mathcal{G}, \mathcal{R}, P) \hookdownarrow \bot$, the last step of our derivation is a read or a write for which we have a $\lnot \tpred{noRacingRd}$ or $\lnot \tpred{noRacingWr}$ for some $i, \mathcal{X}(g)$. Call this step $a_i$. Without loss of generality, we consider the case where $a_i$ is a write. There must also be a write to $g$ by some $j \neq i$, lest, by induction, our derivation does not hit $\bot$. Say the last such write, $a_j$, occurs at step $n$ of the derivation. After the last write $a_j$, it grows monotonically easier to read/write $g$. Hence, there cannot be a $\kw{sync}\, I$ at any step $m > n$ such that $i, j \in I$, since it would be possible for $i$ to read $g$ immediately after the \kw{sync} and, therefore, at all following steps. Hence, there is no $\kw{sync}\ I$ at step $m > n$ on thread $i$ such that $j \in I$ or vice-versa, since that would induce a deadlock, but the hypothesis $(\mathcal{X}_\varnothing, \mathcal{G}, \mathcal{R}, P) \hookdownarrow \bot$ entails that the dynamics run to completion, producing $\bot$---in particular, they do not get stuck. From this fact, we conclude that we can shift our derivation so that $a_j$ and $a_i$ are next to each other. After all, the only steps from other threads that cannot be shifted past $a_i$ are those \textsc{Sync'} where the set includes $i$ and the only steps from other threads that cannot be shifted past $a_j$ (i.e., deleted from the end of the trace) are those \textsc{Sync'} where the set includes $j$. Now, we can transform the piece of the trace before $a_j$, $a_i$ from a $\cdot \hookrightarrow^* \cdot$ into a $\cdot \to^* \cdot$ by Lemma~\ref{lem:678b930f-c61b-4fe8-b686-eaec81d650d5}. Finally, we can append $a_j$, $a_i$ to the result and, separately, $a_i$, $a_j$ to the result to obtain the desired two traces.
\end{proof}

The theorems above characterize executions that run to completion (note the use of $\hookdownarrow$); the next two theorems account for executions that get stuck. We say that a program is \textit{deadlocked} if no step can be taken, but the program is not of the form $\kw{return} \ppar \cdots \ppar \kw{return}$ and, in the checked case, the machine is not in state $\bot$.

\begin{theorem}[Sound Deadlock Detection]
\label{thm:sound-deadlock-detection}
    If any $\cdot \hookrightarrow \cdot$ trace deadlocks from some initial state, then all such traces from that state deadlock. Furthermore, if any $\cdot \hookrightarrow \cdot$ trace deadlocks, then some $\cdot \to \cdot$ trace deadlocks.
\end{theorem}
\begin{proof}
    The first statement is immediate from confluence, noting that, definitionally, the machine cannot step from a deadlocked state. The second statement is immediate from Lemma~\ref{lem:678b930f-c61b-4fe8-b686-eaec81d650d5}.
\end{proof}

\begin{theorem}[Complete Deadlock Detection]
\label{thm:complete-deadlock-detection}
    If some $\cdot \to \cdot$ trace deadlocks, then, for all initial states, all $\cdot \hookrightarrow \cdot$ traces deadlock or reach error state $\bot$.
\end{theorem}
\begin{proof}
    That, for each initial state $\mathcal{X}$, some $\cdot \hookrightarrow \cdot$ trace deadlocks or reaches error state $\bot$ is immediate from Lemma~\ref{lem:dde154e4-0a35-4e5f-998a-804455feda48}. Furthermore, again by confluence, all such traces must deadlock.
\end{proof}

One subtlety of the above theorems is that, if the unchecked semantics both races and deadlocks, the checked semantics might detect one error or the other, though it always detects the same error once we fix the initial state, $\mathcal{X}_\varnothing$ in the implementation since we want the true positives property. We now prove a few more lemmas characterizing $\cdot \hookrightarrow \cdot$ and $\cdot \to \cdot$ to build up to a proof of soundness and completeness for our equivalence checker.

The unchecked dynamics commutes with application of a map to shared memory and the registers. Together Lemmas \ref{lem:678b930f-c61b-4fe8-b686-eaec81d650d5} and \ref{lem:561714cc-cdc8-4e9e-9caf-8ee65e863b7e} (below) should be thought of as a ``sound abstraction'' result: they say that running the checked semantics on a symbolic input tells us everything there is to know about running the unchecked semantics on concrete inputs, as we can take $\varphi : \mathbb{V} \to \mathbb{R}$ in Lemma~\ref{lem:561714cc-cdc8-4e9e-9caf-8ee65e863b7e}:
\begin{lemma}
\label{lem:561714cc-cdc8-4e9e-9caf-8ee65e863b7e}
    For all maps $\varphi$, if $(\mathcal{G}, \mathcal{R}, P) \to^* (\mathcal{G}', \mathcal{R}', P')$ then
    $(\varphi(\mathcal{G}), \varphi(\mathcal{R}), P) \to^* (\varphi(\mathcal{G}'), \varphi(\mathcal{R}'), P')$.
\end{lemma}
\begin{proof}
    By induction on $(\mathcal{G}, \mathcal{R}, P) \to^* (\mathcal{G}', \mathcal{R}', P')$.
\end{proof}

\begin{lemma}
\label{lem:8c00df71-3ceb-42ec-8ddc-bc30c77fc490}
    If $(\mathcal{G}, \mathcal{R}, P) \downarrow (\mathcal{G}_1, \mathcal{R}_1)$ and $(\mathcal{G}, \mathcal{R}, P) \downarrow (\mathcal{G}_2, \mathcal{R}_2)$, then $(\mathcal{G}_1, \mathcal{R}_1) = (\mathcal{G}_2, \mathcal{R}_2)$ or, for all $\mathcal{X}$, $(\mathcal{X}, \mathcal{G}, \mathcal{R}, P) \hookdownarrow \bot$.
\end{lemma}
\begin{proof}
    By Lemma~\ref{lem:dde154e4-0a35-4e5f-998a-804455feda48}, either $(\mathcal{X}, \mathcal{G}, \mathcal{R}, P) \hookrightarrow^* \bot$ and we are done, or we have $(\mathcal{X}, \mathcal{G}, \mathcal{R}, P) \hookrightarrow^* (\mathcal{X}_1, \mathcal{G}_1, \mathcal{R}_1, \kw{return} \ppar \cdots \ppar \kw{return})$ and $(\mathcal{X}, \mathcal{G}, \mathcal{R}, P) \hookrightarrow^* (\mathcal{X}_2, \mathcal{G}_2, \mathcal{R}_2, \kw{return} \ppar \cdots \ppar \kw{return})$ for some $\mathcal{X}_1, \mathcal{X}_2$. But, in the latter case, confluence yields a contradiction if $(\mathcal{G}_1, \mathcal{R}_1) \neq (\mathcal{G}_2, \mathcal{R}_2)$.
\end{proof}

Letting $\tpred{Eq}$ denote our procedure for deciding equality between symbolic terms, we are finally ready to state soundness and completeness of equivalence checking. Soundness says that, if \tpred{Eq} decides that the outputs of the checked dynamics on two programs are equal on appropriate symbolic inputs, then the outputs of the unchecked dynamics on the two programs are equal for all real inputs (since we can write any real inputs as symbolic inputs under a substitution $\varphi$):
\begin{theorem}[Soundness of Equivalence Checking]
\label{thm:sound-equiv-checking}
For all $P,Q : \lang$, contexts $\mathcal{G}, \mathcal{R}$ over $\mathbb{V}$, and maps $\varphi : (\mathbb{V}, \eop) \to (\mathbb{R}, \eop)$, if $(\mathcal{X}_\varnothing, \mathcal{G}, \mathcal{R}, P) \hookdownarrow (\mathcal{G}_P, \mathcal{R}_P)$, $(\mathcal{X}_\varnothing, \mathcal{G}, \mathcal{R}, Q) \hookdownarrow (\mathcal{G}_Q, \mathcal{R}_Q)$, $\tpred{Eq}(\mathcal{G}_P, \mathcal{G}_Q)$, $\tpred{Eq}(\mathcal{R}_P, \mathcal{R}_Q)$, $(\varphi(\mathcal{G}), \varphi(\mathcal{R}), P) \downarrow (\mathcal{G}_P', \mathcal{R}_P')$, and $(\varphi(\mathcal{G}), \varphi(\mathcal{R}), Q) \downarrow (\mathcal{G}_Q', \mathcal{R}_Q')$, then $\mathcal{G}_P' = \mathcal{G}_Q'$ and $\mathcal{R}_P' = \mathcal{R}_Q'$.
\end{theorem}
\begin{proof}
    \emergencystretch=2em
    By Lemmas \ref{lem:678b930f-c61b-4fe8-b686-eaec81d650d5} and \ref{lem:561714cc-cdc8-4e9e-9caf-8ee65e863b7e}, $(\varphi(\mathcal{G}), \varphi(\mathcal{R}), P) \downarrow (\varphi(\mathcal{G}_P), \varphi(\mathcal{R}_P))$ and $(\varphi(\mathcal{G}), \varphi(\mathcal{R}), Q) \downarrow (\varphi(\mathcal{G}_Q), \varphi(\mathcal{R}_Q))$. By Lemma~\ref{lem:8c00df71-3ceb-42ec-8ddc-bc30c77fc490}, either $\varphi(\mathcal{G}_P) = \mathcal{G}_P'$ and $\varphi(\mathcal{R}_P) = \mathcal{R}_P'$ or $(\mathcal{X}_\varnothing, \varphi(\mathcal{G}), \varphi(\mathcal{R}), P) \hookdownarrow \bot$ and thus, by Lemma~\ref{lem:1b7311f3-3b17-43e3-ba53-4c81684cd977}, $(\mathcal{X}_\varnothing, \mathcal{G}, \mathcal{R}, P) \hookdownarrow \bot$, a contradiction. Hence, $\varphi(\mathcal{G}_P) = \mathcal{G}_P'$ and $\varphi(\mathcal{R}_P) = \mathcal{R}_P'$. By the same argument, $\varphi(\mathcal{G}_Q) = \mathcal{G}_Q'$ and $\varphi(\mathcal{R}_Q) = \mathcal{R}_Q'$. Hence, if \tpred{Eq} indeed decides mathematical equality of the symbolic terms (see Section~\ref{sec:decpro}), we have $\mathcal{G}_P' = \varphi(\mathcal{G}_P) = \varphi(\mathcal{G}_Q) = \mathcal{G}_Q'$ and $\mathcal{R}_P' = \varphi(\mathcal{R}_P) = \varphi(\mathcal{R}_Q) = \mathcal{R}_Q'$, as desired.
\end{proof}

Conversely, completeness says that, if \tpred{Eq} decides the outputs of the checked dynamics on two programs are not equal on symbolic inputs, then the outputs of the unchecked dynamics on the two programs are not equal for some real input (namely the symbolic input under substitution $\varphi$):
\begin{theorem}[Completeness of Equivalence Checking]
\label{thm:complete-equiv-checking} 
For all $P,Q : \lang$ and contexts $\mathcal{G}, \mathcal{R}$ over $\mathbb{V}$, if $(\mathcal{G}, \mathcal{R}, \mathcal{X}_\varnothing, P) \hookdownarrow (\mathcal{G}_P, \mathcal{R}_P)$, $(\mathcal{G}, \mathcal{R}, \mathcal{X}_\varnothing, Q) \hookdownarrow (\mathcal{G}_Q, \mathcal{R}_Q)$, and $\lnot \tpred{Eq}(\mathcal{G}_P, \mathcal{G}_Q)$ or $\lnot \tpred{Eq}(\mathcal{R}_P, \mathcal{R}_Q)$, there exists a map $\varphi : (\mathbb{V}, \eop) \to (\mathbb{R}, \eop)$ such that, if $(\varphi(\mathcal{G}), \varphi(\mathcal{R}), P) \downarrow (\mathcal{G}_P', \mathcal{R}_P')$ and $(\varphi(\mathcal{G}), \varphi(\mathcal{R}), Q) \downarrow (\mathcal{G}_Q', \mathcal{R}_Q')$, then $\mathcal{G}_P' \neq \mathcal{G}_Q'$ or $\mathcal{R}_P' \neq \mathcal{R}_Q'$.
\end{theorem}
\begin{proof}
    Without loss of generality, $\lnot \tpred{Eq}(\mathcal{G}_P, \mathcal{G}_Q)$; the cases are symmetric. As we shall show in Section~\ref{sec:decpro}, $\forall a, b : \mathbb{V}$ such that $a \neq b$ there exists a map $\varphi : (\mathbb{V}, \eop) \to (\mathbb{R}, \eop)$ such that $\varphi(a) \neq \varphi(b)$. From this fact, we can obtain $\varphi$ such that $\varphi(\mathcal{G}_P) \neq \varphi(\mathcal{G}_Q)$. By the same argument that we used in the proof of Theorem~\ref{thm:sound-equiv-checking}, if $(\varphi(\mathcal{G}), \varphi(\mathcal{R}), P) \downarrow (\mathcal{G}_P', \mathcal{R}_P')$ and $(\varphi(\mathcal{G}), \varphi(\mathcal{R}), Q) \downarrow (\mathcal{G}_Q', \mathcal{R}_Q')$, then $\varphi(\mathcal{G}_P) = \mathcal{G}_P'$ and $\varphi(\mathcal{G}_Q) = \mathcal{G}_Q'$. If $\mathcal{G}_P' = \mathcal{G}_Q'$, then $\varphi(\mathcal{G}_P) = \mathcal{G}_P' = \mathcal{G}_Q' = \varphi(\mathcal{G}_Q)$, a contradiction. Hence, $\mathcal{G}_P' \neq \mathcal{G}_Q'$ as desired.
\end{proof}

In fact, we only care about equality (or the lack thereof) on a subset of $\mathcal{G}_P, \mathcal{G}_Q$ and $\mathcal{R}_P, \mathcal{R}_Q$, namely the designated program outputs. Fortunately, the same steps prove the desired versions of Theorems~\ref{thm:sound-equiv-checking} and~\ref{thm:complete-equiv-checking}, though their statements become ever more unwieldy.

\subsection{Decision Procedure}
\label{sec:decpro}
Thus far in the formalism, we have assumed, without loss of generality, a single binary operator $\eop$. In the decision procedure, we support expressions involving additions, multiplications, and exponentiations over reals and, in this section, we show that it is possible to decide equalities of the form $f_1(x_1, \ldots, x_n) = f_2(x_1, \ldots, x_n)$ where $f_1$ and $f_2$ involve such expressions. 
Although we do not give a formal account of other functions such as $\max$, the implementation supports them. In principle, $\min$ and $\max$ could be handled by case splits without affecting decidability. In the ML kernels we verify, $\max$ and $\min$ arise only in restricted patterns, such as the max-subtraction of softmax, so we are able to sufficiently normalize them and do not split. We are not aware of a proof of decidability for our desired equalities, though there is work on related problems, such as on Tarski's exponential function problem \cite{realexp}. Closest to our result is Lemma 16 of \citet{li-wu-itcs26} (the basis of {\sc Mirage}'s verifier~\cite{mirage}), a univariate analogue that additionally permits ratios of polynomials in the exponents. However, its proof invokes the Lindemann--Weierstrass theorem and, consequently, applies only when the coefficients involved are algebraic numbers. Our proof is direct, avoids heavy number-theoretic machinery, and handles arbitrary real coefficients.  

If there were no exponentiations, decidability would be immediate. Because $f_1 = f_2$ iff $f_1 - f_2$ is identically zero, and $f_1 - f_2$ can be rewritten as a sum of monomials and such a sum is identically zero iff the coefficients of all the monomials are zero. We show that even with exponentiations, identities $f_1 = f_2$ are decidable.


Without loss of generality, we assume that  the polynomial in the exponent of terms $p(x)e^{h(x)}$ has a zero constant, i.e., $h(0)=0$. If $h(0)\ne 0$ then we can rewrite $p(x)e^{h(x)}$ as $p'(x)e^{h'(x)}$ where $p'(x)=p(x)e^{h(0)}$ and $h'(x)=h(x)-h(0)$. 


\begin{theorem}
    Let $m\geq 1$ be an integer. Suppose that $\mathbb{R}[x]$ polynomials $p_1(x),\ldots,p_m(x)$ and $h_1(x),\ldots,h_m(x)$ satisfy $\forall i\ne j. h_i(x)\ne h_j(x)$ and $\sum_{i=1}^m p_i(x)e^{h_i(x)}=0$ then it must be that  $p_1(x)=p_2(x)=\ldots=p_m(x)=0$.   
\end{theorem}
\begin{proof}
    Proof by induction on $m$. For the base case with $m=1$,
    we have $p_1(x)e^{h_1(x)}=0$. Since $e^{h_1(a)}>0$ for all $a$, it must be that $\forall a. p_1(a)=0$ and hence $p_1(x)=0$.

    For the induction case, assume that the result holds for $(m-1)$. Suppose WLOG that the leading coefficient of $h_m(x)$ is the largest among the leading coefficients of the $h_i(x)$'s. Then the leading coefficient of $h_i(x)-h_m(x)$ is negative for all $i<m$. This implies, $\forall i<m. \lim_{x \to \infty} e^{h_i(x)-h_m(x)}=0$.

    We are given, $p_1(x)e^{h_1(x)-h_m(x)}+p_2(x)e^{h_2(x)-h_m(x)}+\ldots+p_{m-1}(x)e^{h_{m-1}(x)-h_m(x)}+p_m(x)=0$.
    Taking limits as $x$ approaches $\infty$, we have:
    \[
    \left(\sum_{i<m}\lim_{x\to\infty} p_i(x)e^{h_i(x)-h_m(x)}\right) + \lim_{x\to\infty} p_m(x)=0
    \]
    Therefore, $\lim_{x\to\infty} p_m(x)=0$. But for a non-zero polynomial $p$, $\lim_{x\to\infty} p(x)$ is either $\infty$ if the leading coefficient of $p$ is positive and $-\infty$ if negative. Hence, $p_m(x)=0$.
\end{proof}

Hence, in the uni-variate case, identity checking with exponentiations reduces to identity checking without exponentiations, which is decidable as discussed earlier. The above theorem for uni-variates generalizes to multi-variate polynomials.

\begin{corollary}
\label{cor:multivariate-identity}
Let $m\geq 1$ be an integer. Suppose that $\mathbb{R}[\bar{x}]$ multivariate polynomials $p_1(\bar{x}),\ldots,p_m(\bar{x})$ and $h_1(\bar{x}),\ldots,h_m(\bar{x})$ satisfy $\forall i\ne j. h_i(\bar{x})\ne h_j(\bar{x})$ and $\sum_{i=1}^m p_i(\bar{x})e^{h_i(\bar{x})}=0$ then it must be that  $p_1(\bar{x})=p_2(\bar{x})=\ldots=p_m(\bar{x})=0$. 
\end{corollary}
\begin{proof}
    Consider the substitution $x_1=\alpha_1y$, $x_2=\alpha_2y$, $\ldots$, $x_n=\alpha_ny$ where each $\alpha_i\in\mathbb{R}$ are chosen independently at random. Making this substitution, we get a univariate functional identity of the form 
    \begin{equation}
    \label{eq:corrolary}
    \hat{p}_1(y)e^{\hat{h}_1(y)}+\hat{p}_2(y)e^{\hat{h}_2(y)}+\ldots+\hat{p}_m(y)e^{\hat{h}_m(y)}=0
    \end{equation}
    where for each $i\in\{1,\ldots,m\}$, $\hat{p}_i(y)=p_i(\alpha_1y,\alpha_2y,\ldots,\alpha_ny)$
    and $\hat{h}_i(y)=h_i(\alpha_1y,\alpha_2y,\ldots,\alpha_ny)$.

    From our convention, $\hat{h}_i(0)=h_i(0,0,\ldots,0)\ne 0$.
    For $i\ne j$, consider \[\hat{h}_i(y)-\hat{h}_j(y)=h_i(\alpha_1y,\ldots,\alpha_ny)-h_j(\alpha_1y,\ldots,\alpha_ny)\]
    Let the degree of $h_i(\bar{x})-h_j(\bar{x})$ be $d$ and the largest degree homogeneous component of $h_i(\bar{x})-h_j(\bar{x})$ be $g_d(\bar{x})\ne 0$. Then, $\hat{h}_i(y)-\hat{h}_j(y)$ is a polynomial of degree at most $d$ and the coefficient of $y^d$ in $\hat{h}_i(y)-\hat{h}_j(y)$ is $g_d(\alpha_1,\alpha_2,\ldots,\alpha_n)$. By Schwartz Zippel Lemma,  $g_d(\alpha_1,\alpha_2,\ldots,\alpha_n)\ne 0$ almost surely and so almost surely (over the random choice of $\bar{\alpha}=(\alpha_1,\ldots,\alpha_n)$) we have $\forall i\ne j: \hat{h}_i(y)\ne \hat{h}_j(y)$.
    Applying the theorem to Equation~\ref{eq:corrolary}, we get almost surely (over the random choice of $\bar{\alpha}$, $\hat{p}_1(y)=\ldots=\hat{p}_m(y)=0$). But $\hat{p}_i(y)=p_i(\alpha_1y,\ldots,\alpha_my)$ and this must imply that $p_i(\bar{x})=0$ (via Schwartz Zippel Lemma applied to the largest homogeneous component of $p_i(\bar{x})$). Thus, $\forall 1\leq i\leq m. p_i(x)=0$.
\end{proof}

Note that the theorem and corollary require the exponents $h_i$ to be polynomials. Nested exponentials, as would arise, for example, if one softmax were composed with another, are not covered by our decidability proof. Our implementation simplifies nested exponentials and ratios of functions in exponents if they arise, but it may not be complete for such expressions.

Given our interest in proving completeness of the overall equivalence checking procedure, we do not focus on the computational complexity of the decision procedure and leave it for future work. Our implementation (Section~\ref{sec:impl}) is able to handle ML kernels that arise in practice (Section~\ref{sec:evaluation}). Finally, to make good on our claim in the proof of completeness for equivalence checking that $\forall a, b : \mathbb{V}$ such that $a \neq b$ there exists a map $\varphi : (\mathbb{V}, \eop) \to (\mathbb{R}, \eop)$ (more precisely $(\mathbb{V}, +, \cdot, \exp) \to (\mathbb{R}, +, \cdot, \exp)$) such that $\varphi(a) \neq \varphi(b)$, we remark that $a, b$ can be viewed as polynomials in $\mathbb{R}[\bar{x}]$ for some $\bar{x}$. By contraposition of the above theorem, recalling that each $p_i$ can be written as a sum of monomials, if $a \neq b$ (that is, $a - b$ is not formally $0$), they must be unequal on some input $\bar{r}$. Then,
\begin{align*}
    \varphi(x) = \begin{cases}
        r_i &\text{if } x = x_i \text{ for some } x_i \in \bar{x} \\
        x &\text{otherwise}
    \end{cases}
\end{align*}
can be extended in the obvious way to the desired map $(\mathbb{V}, +, \cdot, \exp) \to (\mathbb{R}, +, \cdot, \exp)$.

\section{Implementation}
\label{sec:impl}
We implement our techniques in a proof-of-concept tool, \tool, written in Rust.\footnote{\url{https://github.com/willtunnels/volta}} \tool operates on CTAs written in PTX ISA 9.1\footnote{\url{https://docs.nvidia.com/cuda/parallel-thread-execution/}}. In addition to the PTX program, it accepts specification of parameters such as the grid dimensions and the locations and sizes of input tensors as well as the output tensors that should be compared for equivalence. \tool is comprised of two main components: a simulator that performs symbolic execution to generate verification conditions (VCs), and a decision procedure that resolves them. For a program where each thread runs at most $N$ statements, the simulator's runtime is linear in $N$, whereas the decision procedure exhibits superlinear complexity in $N$. The rest of this section will relay the key details of symbolic execution and the decision procedure, though it does not exhaustively describe their implementations.

Before symbolic execution, the input PTX is lowered, resolving variable, parameter, and register names, as well as labels. During symbolic execution, for each shared or global memory cell that has been written, we track an expression and the value of $\mathcal{X}$ (see Section~\ref{sec:symex}) at that cell, where $\mathscr{P}(\text{TID})$ is implemented via a fixed-width bitset. In fact, memory cells additionally track the width of the last store and require (to a first approximation) that reads are of the same width---it is not clear what it means to read half of a real valued symbolic expression. Each expression is passed around as a 4-byte index into a bump allocator holding a 16-byte representation of the expression. Constant folding, including simplification of $\min(\infty, x)$, $\min(x, \infty)$, $\max(-\infty, x)$, and $\max(x, -\infty)$, is performed eagerly as expressions are constructed, with floating-point constants represented via GMP's~\cite{gmp} arbitrary precision rationals. Then, it is a trivial matter to resolve branches and memory accesses as they occur, should they be fully statically determined, or else to return an error.

Threads are scheduled in a round-robin fashion, with each selected thread being run until it blocks at a barrier or exits. If all threads are blocked, the scheduler checks whether any barriers can be released because all threads that are expected to reach them have done so. If not, a deadlock is reported---a strategy justified by Theorems~\ref{thm:sound-deadlock-detection} and~\ref{thm:complete-deadlock-detection}. At any point, a race might be detected via $\mathcal{X}$, it might be decided that the program is not a structured-CTA because it attempts to branch or index memory using a symbolic value, or some other violation might be detected such as an out-of-bounds read. If the symbolic execution succeeds, the verification conditions (VCs) corresponding to the output tensors are given to a decision procedure. In particular, for each pair of corresponding output tensor elements $p_1$ and $p_2$, the decision procedure checks whether $p_1 - p_2$ is identically zero.

\tool currently assumes that CTAs behave uniformly and simply checks CTA 0 of the reference against CTA 0 of the optimized program. If the optimizations are intra-CTA, checking multiple CTAs is embarrassingly parallel, but it might become intractably expensive for real, multi-billion parameter models. It is worth noting that, while the number of CTAs grows, the per-CTA tensor size does not.

The decision procedure consists of a canonicalizer producing a ratio of polynomials (rational function), each polynomial consisting of a sorted list of terms along with their coefficients, each term consisting of a sorted list of factors along with their exponents, optionally multiplied by $e$ to some (non-zero) polynomial, each factor consisting of a symbolic input, an uninterpreted function of rational functions, or a min or max of rational functions. Min and max factors store their arguments sorted and deduplicated, and nested min and, respectively, max factors are flattened. Everything is hash-consed, providing an arbitrary but stable sort order. Rational functions are not reduced to lowest terms; equality of two canonical ratios is decided by a monomial-quotient test that handles softmax-style rescalings, or, in general, by cross-multiplication, which reduces the question to equality of canonical polynomials. The results of canonicalization are memoized/cached based on the generation time indices that represent expressions, but only if an expression is referenced by multiple other expressions, where reference counts are determined by a walk of the generation phase expression store (the bump allocator) before running the decision procedure. This check significantly reduces the size of the memoized data, e.g., in the presence of accumulation chains, and the memoized data is not incrementally collected.

The content of Corollary~\ref{cor:multivariate-identity} is that the above procedure is complete for the subclass of expressions $\sum_{i=1}^m p_i(\bar{x})e^{h_i(\bar{x})}$. It says such functions are determined by their polynomial coefficients (which are themselves determined by their monomial coefficients). Canonicalization requires distributing multiplication over addition to find the coefficient of each monomial term. This operation, in principle, may cause exponential blowup. Since machine learning workloads do not typically have computations with high multiplicative depth, this blowup does not happen in practice.

It is worth noting that \tool's implementation is currently single-threaded. One could try to parallelize VC generation by simulating the CUDA parallel machine with a parallel machine. The decision procedure would be a more interesting task because of canonical form memoization, out of which the implementation gets quite a bit of mileage, as we shall see (Section~\ref{sec:equiv-check-baseline}). We leave parallelization to future work.

Finally, we give an inventory of the PTX operations \tool currently supports. It supports collective operations \texttt{bar.sync}, \texttt{bar.warp.sync}, \texttt{shfl.sync}, \texttt{activemask}, \texttt{ldmatrix}, \texttt{mma.sync}, and \texttt{wmma}. It can symbolically compute addition, multiplication, division, reciprocal, fused multiply-add, and exponentiation along with max and min, float-width conversions (modeled as the identity), and floating-point saturation and ReLU clamps (modeled via min/max). Some additional operations can be performed on statically known values.
We do not support square root or trigonometric functions. These operations are tricky because they place requirements on their arguments, e.g., square root requires a positive argument. If the set of operations appearing in expressions grows sufficiently complex, an SMT solver might be necessary.

\section{Evaluation}
\label{sec:evaluation}

There are three main ways to create efficient GPU kernels for machine learning workloads: manual optimization (Section~\ref{sec:human}), generation by LLMs (Section~\ref{sec:model}), and compilation by domain-specific compilers (Section~\ref{sec:compiler}). Since our equivalence checker operates as a black box and makes no assumptions about the optimization process, it can verify the correctness of optimized kernels produced by all three approaches. We evaluate on  reductions, matrix multiplications, convolutions, and various attention mechanisms. We report the time to generate the verification conditions (VCs), which includes symbolically executing both kernels (``VC Gen''), and the time to solve them (``VC Time''), measured on a machine with a 128-core 2.25 GHz AMD EPYC 7742 processor and 995 GB of RAM.
We show that \tool scales to realistic machine learning workloads and effectively identifies correctness issues when they exist. In particular, \tool detects data races that have been previously reported in open-source GitHub repositories (Section~\ref{sec:dr}).

\subsection{Human-generated Kernels}
\label{sec:human}
\subsubsection{Reduction}
``Optimizing Parallel Reduction in CUDA'' by Harris~\cite{reduction} is a well-cited tutorial that introduces a sequence of progressively optimized reduction kernels. It begins with a baseline implementation, Red-1, and presents seven versions, each achieving higher performance than the previous one. Each CTA reduces 128 values using up to 128 threads. The first optimization, Red-2, removes divergent branching and achieves more than a $2\times$ speedup. The next version, Red-3, mitigates shared-memory bank conflicts for another $2\times$ improvement. Red-4, which further reduces the number of idle threads, yields an additional reported speedup of $1.78\times$. The subsequent optimizations rely on warp-synchronous execution, which was later deprecated by NVIDIA. \tool detects data races in these versions, Red-5, Red-6, and Red-7, and correctly rejects them, and \tool proves that the other kernels are equivalent to Red-1.

Table~\ref{tab:red} reports the running time of \tool and other key statistics, such as the number of CTA-wide synchronization operations and PTX instructions, for the optimized kernels. Red-1 through Red-4 were checked for equivalence with Red-1; for the racy versions, we report the time for symbolic execution to detect the race. ``\#Block Sync'' represents the total number of {\tt bar.sync} PTX instructions executed across all threads.
Finally, verifying the equivalence of a kernel against itself, here Red-1 vs.\ Red-1, serves as a sanity check, confirming that the reference implementation is free of synchronization errors.

\begin{table}[h!]
\caption{Time to symbolically execute and check verification conditions (VCs) for kernels from NVIDIA's reduction tutorial~\cite{reduction}. Each kernel runs as a single CTA of 128 threads; Red-1--Red-4 reduce 128 elements and were checked for equivalence with Red-1. \tool rejects the racy Red-5--Red-7 during symbolic execution, before any VCs are generated, so their VC Gen entries report the time to detect the race.}
\label{tab:red}
\centering
{\footnotesize\begin{tabular}{lcccc}
\toprule
\textbf{Kernel} & \textbf{VC Gen (s)} & \textbf{VC Time (s)} & \textbf{\#Block Sync} & \textbf{PTX LOC} \\
\midrule
Red-1              & 0.0038 & 0.0005  &  1k    &  76  \\
Red-2              & 0.0023 & 0.0004  &  1k    &  79\\
Red-3              & 0.0022 & 0.0004  &  1k    &  73\\
Red-4              & 0.0022 & 0.0004  &  896    &  75\\
Red-5              & 0.0011 & ---  &  ---    &  156\\
Red-6              & 0.0007 & ---  &  ---    &  104\\
Red-7              & 0.0010 & ---  &  ---    &  98\\
\bottomrule
\end{tabular}
}
\end{table}

\subsubsection{MatMul}
``How to Optimize a CUDA Matmul Kernel for cuBLAS-like Performance'' by  Boehm is a widely referenced tutorial in the GPU community~\cite{boehm} presenting SGEMM (single-precision general matrix multiplication) implementation techniques. Similar to the reduction tutorial, it presents a series of kernels, each achieving progressively higher performance: starting from 1\% of the performance of NVIDIA's cuBLAS (proprietary and not open source) and reaching 95\%. Each CTA computes a $64\times64$ tile of the output matrix using up to 512 threads. Starting from a baseline implementation, MatMul-1, the subsequent versions, MatMul-2 through MatMul-7, apply a sequence of standard optimizations, including coalesced global-memory accesses, use of shared-memory for caching, reducing bank conflicts, multiple levels of tiling (including warp tiling), vectorized memory operations, and auto-tuning to determine the optimal sub-tile sizes for intermediate computations. As with the reduction kernels, Table~\ref{tab:boehm} reports the running time, the number of CTA-wide synchronization operations, and the number of PTX instructions for each optimized variant. All kernels were checked for equivalence with MatMul-1. In contrast to the reduction tutorial, all MatMul kernels pass the equivalence check without data races.

\begin{table}[h!]
\caption{Symbolic execution and VC checking time for kernels from matrix multiplication tutorial~\cite{boehm}. All kernels were checked for equivalence with MatMul-1. Each CTA computes a $64\times 64$ tile of $C = \alpha AB + \beta C$ with $M{=}N{=}K{=}4096$; MatMul-1 and MatMul-2 use 512 threads per CTA, MatMul-3 through MatMul-5 use 128, and MatMul-6 and MatMul-7 use 256.}
\label{tab:boehm}
\centering
{\footnotesize\begin{tabular}{lcccc}
\toprule
\textbf{Kernel} & \textbf{VC Gen (s)} & \textbf{VC Time (s)} & \textbf{\#Block Sync} & \textbf{PTX LOC} \\
\midrule
MatMul-1              & 45.4 & 89.4  &  524k    & 185   \\
MatMul-2              & 38.5 & 90.2  & 524k    &  269 \\
MatMul-3              & 30.4 & 90.5  &  131k    &  325\\
MatMul-4              & 30.7 & 91.7  &  131k    &  323\\
MatMul-5              & 30.6 & 92.8  &  131k    &  320\\
MatMul-6              & 34.7 & 94.0  &  262k    &  307\\
MatMul-7              & 34.5 & 93.5  &  262k    &  307\\
\bottomrule
\end{tabular}
}
\end{table}
\subsubsection{FlashAttention}
The performance of LLM inference is often bottlenecked by the attention operation. Consequently, optimizing attention has become a central topic in the performance engineering of LLMs. We start from a naive reference implementation, Attention,  that uses a single thread to compute the full attention operation, i.e., $\mathit{softmax}(QK^T / \sqrt{d})V$. We evaluate the setup where each CTA computes an attention head with standard matrix dimensions $Q:(16,64)$, $K:(512,64)$, and $V:(512,64)$. The optimized implementations we consider next use 128 threads.

FA1 applies the FlashAttention~\cite{fa} tiling strategy to reduce global-to-shared memory transfers and adopts the online softmax formulation to enable streaming. FA1-TC further accelerates FA1 by leveraging tensor cores. Finally, FA2-TC also uses tensor cores and incorporates the techniques of FlashAttention-2~\cite{fa2}, such as deferring softmax normalization and improving work partitioning among warps. Table~\ref{tab:attn} reports the checker runtime and statistics for these optimized variants, including the number of CTA-level barriers ({\tt bar.sync}), warp-level barriers, PTX instruction counts, and the number of threads. ``\#Warp Sync" represents the total number of warp-level synchronization PTX instructions such as {\tt bar.warp.sync}, {\tt mma.sync}, etc., executed across all threads. Note that kernels in Table~\ref{tab:red}, Table~\ref{tab:boehm}, Attention, and FA1 do not use tensor cores. Thus, they have zero warp-level synchronization operations. All kernels were checked for equivalence with Attention. The Attention row checks the reference against itself---a sanity check as in Table~\ref{tab:red}. Its two sides canonicalize to identical expressions, so their equality is immediate. The FlashAttention rows instead reconcile two different representations of each output element as a rational function---the naive softmax against the online-softmax accumulation---and deciding these equalities dominates their VC Times.

\begin{table}[h!]
\caption{VC generation and checking time for attention kernels. All kernels were checked for equivalence with Attention. Each kernel computes one attention head with $Q:(16,64)$, $K:(512,64)$, and $V:(512,64)$ using a single CTA with the listed number of threads.}
\label{tab:attn}
\centering
{\footnotesize
\begin{tabular}{lcccccc}
\toprule
\textbf{Kernel} & \textbf{VC Gen (s)} & \textbf{VC Time (s)} & \textbf{\#Block Sync} & \textbf{\#Warp Sync} & \textbf{PTX LOC} & \textbf{\#Threads}  \\
\midrule
Attention            & 3.9 & 3.0  & 0   & 0   & 477  & 1     \\
FA1            & 6.7 & 153.8  & 187k   & 0  & 1138  & 128     \\
FA1-TC     & 8.1 & 154.5  & 801k   & 204k   & 1237  & 128    \\
FA2-TC & 4.1 & 154.0  & 210k   & 13k   & 2048  & 128  \\
\bottomrule
\end{tabular}
}
\end{table}

Versions of attention which feature causal masking are commonly used in auto-regressive models. We verify that the efficient ``fused'' implementation of causal masking, wherein we skip any blocks such that all the column indices are more than the row indices, is equivalent to a naive approach implemented via an element-wise product with a constant triangular matrix. See Table~\ref{tab:causal-attn}.

VC checking time for these benchmarks is significantly lower than the VC checking time for the benchmarks in Table~\ref{tab:attn}. The symbolic expressions are smaller due to masking and time is non-linear in expression size. Moreover, the two expressions being compared are actually precisely the same---since the only difference is the method of filtering and this difference evaporates during symbolic evaluation---allowing the implementation to avoid work via the cache.

\begin{table}[h!]
\caption{VC generation and checking time for causal attention kernels with a lower triangular mask. For all kernels, a fused version of the kernel was checked against a naive version of the kernel. The first element of PTX LOC is the for the fused implementation, and second element is for the naive implementation. Dimensions and launch configurations match those of Table~\ref{tab:attn}.}
\label{tab:causal-attn}
\centering
{\footnotesize
\begin{tabular}{lcccccc}
\toprule
\textbf{Kernel} & \textbf{VC Gen (s)} & \textbf{VC Time (s)} & \textbf{\#Block Sync}  & \textbf{\#Warp Sync} & \textbf{PTX LOC} & \textbf{\#Threads} \\
\midrule
Causal-Attention & 3.9 & 0.05  & 0 & 0 & (483, 480) & 1 \\
Causal-FA1 & 9.4 & 0.08 & 187k & 0 & (1141, 1142) & 128 \\
Causal-FA1-TC & 11.8 & 0.13 & 801k & 204k & (1241, 1242) & 128 \\
Causal-FA2-TC & 4.3 & 0.07 & 210k & 13k & (2081, 2088) & 128 \\
\bottomrule
\end{tabular}
}
\end{table}

\subsection{LLM-generated Kernels}
\label{sec:model}
In a recent blog post, Stanford researchers demonstrated how LLMs can generate optimized GPU kernels for 2D convolution~\cite{crfm}. We verify the equivalence of the final LLM-generated kernel after 13 rounds of optimization with a reference implementation. The applied optimizations include transforming the convolution into a matrix multiplication to leverage tensor cores, pre-computing intermediate values to eliminate redundant arithmetic, caching indices in shared-memory, employing software pipelining, reusing address computations, and applying vectorization. Overall, the final kernel achieves 179.9\% of PyTorch performance. In this setup, each CTA uses 256 threads to compute a $128 \times 128$ output tile. Table~\ref{tab:conv} reports \tool's running time and other key statistics. The use of tensor cores significantly increases the code size of the optimized kernel compared to the reference. The reference has no synchronization operations whereas the final version has 6.4k CTA-wide and 102k warp-level barriers.

\begin{table}[h!]
\caption{Symbolic execution and VC checking time for the LLM-generated 2D convolution~\cite{crfm}, checked against a naive reference implementation. In the pairs $(-,-)$, the first element is for the reference implementation, and the second element is for the optimized implementation. The kernels compute a 2D convolution (batch 100, $3\times 224\times 224$ inputs, 96 $11\times 11$ filters, stride 4, padding 2) expressed as an implicit GEMM with $M{=}96$, $N{=}302{,}500$, $K{=}363$; each CTA uses 256 threads and computes a $128\times 128$ output tile.}
\label{tab:conv}
\centering
{\footnotesize
\begin{tabular}{lcccccc}
\toprule
\textbf{Kernel} & \textbf{VC Gen (s)} & \textbf{VC Time (s)} & \textbf{\#Block Sync} & \textbf{\#Warp Sync} & \textbf{PTX LOC} & \textbf{\#Threads}  \\
\midrule
Conv2D & 32.0 & 14.3  & (0, 6.4k)   & (0, 102k)   & (463, 1312)  & (256, 256)     \\
\bottomrule
\end{tabular}
}
\end{table}

Through the course of this verification, we discovered interesting undocumented behaviors. For example, out-of-bounds reads from shared-memory arrays may not cause a crash and may also go undetected by NVIDIA's Compute Sanitizer~\cite{compute-sanitizer}.
The following code segment---which uses two warps (64 threads) to read 48 elements of the shared-memory array {\tt A} into registers {\tt v}---runs without raising any exceptions on the Ampere, Hopper, and Blackwell GPUs we tested on, even though it performs out-of-bounds accesses.
\begin{lstlisting}[language=C, basicstyle=\ttfamily\scriptsize]
__shared__ int A[48];
...
v = A[tid]; // threads 48 to 63 read out-of-bound
if (tid < 48){
  use V
...
\end{lstlisting}
The out-of-bounds reads store zeros in register {\tt v} and do not crash the tests. 
\tool (correctly) catches them and raises exceptions. An appropriate fix is to rewrite the code, as shown below:
\begin{lstlisting}[language=C, basicstyle=\ttfamily\scriptsize]
__shared__ int A[48];
...
if(tid < 48){
  v = A[tid]; // all accesses within bounds
  use V
...
\end{lstlisting}
Although the two programs behave identically on current GPUs, there is no guarantee that future GPU hardware would continue to tolerate out-of-bounds reads to shared memory. The latter code segment is correct for future hardware as well and, hence, preferable. Thus, even though we focus on verifying correctness of optimizations in this paper, we believe \tool can also help improve code quality.

Next, we use Claude Code (Anthropic's LLM-based coding tool) to improve performance through an agentic setup that automatically generates and runs tests for correctness checking~\cite{claudecode}.
In Table~\ref{tab:agent}, we start with a reference implementation for a matrix multiplication, MatMul-1, adapted from~\cite{boehm}, where the CTAs output  $32\times 32$ tiles.  Claude Code is able to generate various kernels where GEMM-1 and GEMM-2 have shared-memory optimizations, and GEMM-3 uses tensor cores. \tool is successfully able to prove the equivalence of all three with the reference implementation.

\begin{table}[h!]
\caption{Symbolic execution and VC checking time for matrix multiplication kernels generated by Claude Code. All kernels were checked for equivalence with MatMul-1. Each CTA uses 512 threads to compute a $32\times 32$ tile of $C = AB$ with $M{=}N{=}K{=}4096$.}
\label{tab:agent}
\centering
{\footnotesize
\begin{tabular}{lcccccc}
\toprule
\textbf{Kernel} & \textbf{VC Gen (s)} & \textbf{VC Time (s)}  &\textbf{\#Block Sync} & \textbf{\#Warp Sync} & \textbf{PTX LOC} & \textbf{\#Threads} \\
\midrule
GEMM-1              & 36.4 & 13.2  &  524k  & 0  & 562  & 512 \\
GEMM-2              & 39.9 & 13.1  & 262k   & 0 &  587 & 512\\
GEMM-3              & 28.4 & 13.1  &  263k  & 98k  &  771 & 512\\
\bottomrule
\end{tabular}
}
\end{table}

\subsection{Compiler-generated Kernels}
\label{sec:compiler}
TileLang~\cite{tilelang} is a compiler that takes as input Python code that uses TileLang's DSL API and generates CUDA kernels under different configuration settings. We use it to produce multiple implementations of matrix multiplication and verify their pairwise equivalence. Any equivalence failure would indicate a correctness bug in the TileLang compiler. Table~\ref{tab:tilelang} shows that the running time of \tool scales gracefully with increasing tile sizes.  Note that these tiles cannot grow unbounded in size as then they would exceed the resource limits that the hardware imposes on each CTA. Hence, Table~\ref{tab:tilelang} evaluates on some common tile sizes, where $A\times B\times C$ indicates a multiplication of a matrix with dimensions $A\times B$ with a matrix of dimensions $B\times C$. Table~\ref{tab:tilelang} also reports detailed statistics for each pair of programs under verification. In all cases, the optimized implementations use tensor cores whereas their reference counterparts do not. The first elements of the pairs in Table~\ref{tab:tilelang} refer to the statistics for the reference implementations and the second elements correspond to the optimized implementations. The optimized implementations call into CUTLASS templates, thus demonstrating \tool's abilities to handle PTX assembly arising from CUTLASS~\cite{cutlass}.

\begin{table}[h!]
\caption{VC generation and checking time for matrix multiplication kernels generated by TileLang~\cite{tilelang}. In each pair $(-,-)$, first element is the reference implementation, and second element is the optimized implementation. All kernels multiply f16 matrices with $M{=}N{=}K{=}4096$; each CTA computes one output tile of the listed size using the listed number of threads.}
\label{tab:tilelang}
\centering
{\footnotesize
\begin{tabular}{lcccccc}
\toprule
\textbf{Tile size} & \textbf{VC Gen (s)} & \textbf{VC Time (s)} & \textbf{\#Block Sync} & \textbf{\#Warp Sync} & \textbf{PTX LOC} & \textbf{\#Threads}  \\
\midrule
$32\times 32\times 32$            & 1.1 & 2.5  & (8k, 8k)   & (0, 32k)   & (322, 455)  & (128, 128)     \\
$64\times 32\times 32$    & 4.5 & 10.9  &  (16k, 16k)   & (0, 115k)   & (531, 537)  & (128, 128)    \\
$64\times 64\times 32$  & 18.6 & 60.4  &  (32k, 32k)   & (0, 393k)   & (632, 842)  & (128, 128)   \\
\bottomrule
\end{tabular}
}
\end{table}

\begin{table}[h!]
\caption{Racy FaialAA GitHub benchmarks; races are also detected by \tool before the fix, and not detected after the fix. BucketPositions, ComputeRange, and ReduceValue run as a single CTA of 64 threads; LayerNorm runs as a single warp (32 threads); GradInput runs with $32\times 4$ threads before the fix and $32\times 1$ after.}
\label{tab:race}
\centering
{\footnotesize\begin{tabular}{lccc}
\toprule
\textbf{Kernel} & \textbf{Library} & \textbf{PTX LOC} & \textbf{Bug Fix Commit}   \\
\midrule
BucketPositions & OpenMM  & 94 & \texttt{9abaa587c} \\
ComputeRange & OpenMM  & 132 & \texttt{65caf22b4} \\
ReduceValue  & OpenMM  & 1190 & \texttt{de666e305} \\
LayerNorm  & Megatron-LM  & 924 & \texttt{4831071c9} \\
GradInput  & Megatron-LM  & 546 & \texttt{f1295380c} \\
\bottomrule
\end{tabular}
}
\end{table}

\subsection{Data Race Detection Baseline}
\label{sec:dr}
FaialAA is the state-of-the-art data race checker for GPU programs~\cite{faialaa}. Its canonical failure case, as described by the authors, is the following pattern:
\begin{lstlisting}[basicstyle=\ttfamily]
1. M[tid] = tid;
2. x = M[tid];
3. M[x] = ...;
\end{lstlisting}
FaialAA cannot track the concrete value of {\tt x} at line 2 and, therefore, models it as non-deterministic. Subsequently, line 3 is reported as a (spurious) data race. Since we use symbolic evaluation, and \texttt{M[x]} is constant (since \texttt{tid} is constant), we don't report an error and declare the write in line 3 correctly as data race free.

FaialAA evaluates on GitHub commits that fix data races.
These programs come from well-used open-source repositories OpenMM\footnote{\url{https://github.com/openmm/openmm}} and NVIDIA's Megatron-LM\footnote{\url{https://github.com/NVIDIA/Megatron-LM}}.
Empirically, on the programs where FaialAA successfully detects data races, \tool succeeds as well, i.e., the data races are caught by the symbolic evaluation. \tool also successfully proves data race freedom for the versions of these kernels after the bug-fixing commits. These programs are shown in Table~\ref{tab:race} for completeness.

\subsection{Equivalence Checking Baseline}
\label{sec:equiv-check-baseline}

To check that our decision procedure yields some benefit, we compare it to the \textsc{Z3} \citep{z3} SMT solver, an ``off-the-shelf'' solution, giving each the VC for a single element of each benchmark's output tensor. Table~\ref{tab:z3-comparison} reports the running time of \tool's decision procedure and of \textsc{Z3}, with and without the exponential-product axiom, $\forall x,y.\; e^x e^y = e^{x+y}$, on the benchmarks of Section~\ref{sec:evaluation}. The \textsc{Z3} times and the \tool ``First'' times are for verifying the VC of a single element of the output tensor in isolation. The \tool ``Median'' column reports the median per-element time from the full-footprint runs of Section~\ref{sec:evaluation}, in which \tool checks every output element.

On the 18 benchmarks whose VCs contain no exponentials (the reduction, matrix multiplication, and convolution kernels), \textsc{Z3} succeeds. On the eight attention benchmarks, whose VCs contain exponentials by way of softmax, \textsc{Z3} succeeds only on those VCs which are trivial after symbolic execution (exactly the form $p - p$ for some $p$). Adding the axiom does not help---\textsc{Z3} simply runs longer before giving up. The gap between \tool's first and median solve times has two sources. First, the initial check of a run performs a one-time analysis of both kernels' full expression graphs to determine expression reference counts (see Section~\ref{sec:impl}); for the matrix multiplications, whose elements share little structure, this is essentially the entire gap. Second, an element checked after its neighbors reuses cached canonical forms for structure it shares with them; for the attention kernels, a row's softmax normalization chain is canonicalized by the row's first element and reused by the remaining 63 elements of the row. \textsc{Z3}, by contrast, solves each element's VC as an independent query at unchanged cost. Hence, \tool's first solve time is often significantly higher than \textsc{Z3}'s, but its median solve time is significantly lower.

\begin{table}[h!]
\caption{Per-element VC times using \tool's decision procedure and \textsc{Z3}, without and with the axiom $\forall x,y.\; e^x e^y = e^{x+y}$. ``Volta First'' solves the VC of one output tensor element in isolation; ``Volta Median'' is the median per-element time when solving the VCs of all output elements in one pass, which amortizes one-time setup and shared canonicalization work across the elements. For the reductions, whose output is a single element, the two \tool columns measure the same work in different runs. The axiom is relevant only when a VC contains exponentials, so it is omitted for the other benchmarks. ``Unknown'' indicates that \textsc{Z3} returned {\tt unknown} after the parenthesized time. Both \tool and \textsc{Z3} version 4.8.12 were given a single thread and a 10 minute timeout, running on a 128-core 2.25 GHz AMD EPYC 7742 processor with 995 GB of RAM.}
\label{tab:z3-comparison}
\centering
{\footnotesize
\begin{tabular}{lcccc}
\toprule
\textbf{Kernel} & \textbf{Volta First (ms)} & \textbf{Volta Median (ms)} & \textbf{Z3 (ms)} & \textbf{Z3 + Axiom (ms)} \\
\midrule
Red-1 & $<$ 1 & $<$ 1 & 3.6 & --- \\
Red-2 & $<$ 1 & $<$ 1 & 3.6 & --- \\
Red-3 & $<$ 1 & $<$ 1 & 3.6 & --- \\
Red-4 & $<$ 1 & $<$ 1 & 3.6 & --- \\
MatMul-1 & 435 & 18 & 45 & --- \\
MatMul-2 & 358 & 17 & 47 & --- \\
MatMul-3 & 385 & 17 & 47 & --- \\
MatMul-4 & 383 & 17 & 49 & --- \\
MatMul-5 & 383 & 17 & 48 & --- \\
MatMul-6 & 385 & 18 & 50 & --- \\
MatMul-7 & 388 & 18 & 47 & --- \\
Conv2D & 313 & $<$ 1 & 5.1 & --- \\
GEMM-1 & 165 & 10 & 50 & --- \\
GEMM-2 & 163 & 10 & 49 & --- \\
GEMM-3 & 162 & 10 & 47 & --- \\
$32 \times 32 \times 32$ & 16 & 1.9 & 9.3 & --- \\
$64 \times 32 \times 32$ & 46 & 4.4 & 17 & --- \\
$64 \times 64 \times 32$ & 217 & 12 & 45 & --- \\
\hline\hline
Attention & 141 & 1.3 & 395 & 388 \\
FA1 & 6385 & 90 & Unknown (3.0 s) & Unknown (319 s) \\
FA1-TC & 6481 & 88 & Unknown (3.0 s) & Unknown (319 s) \\
FA2-TC & 6382 & 89 & Unknown (3.0 s) & Unknown (317 s) \\
Causal-Attention & 11 & $<$ 1 & 3.7 & 3.6 \\
Causal-FA1 & 33 & $<$ 1 & 3.8 & 3.8 \\
Causal-FA1-TC & 98 & $<$ 1 & 3.8 & 3.8 \\
Causal-FA2-TC & 20 & $<$ 1 & 3.9 & 3.8 \\
\bottomrule
\end{tabular}
}
\end{table}

\section{Related Work}
\label{sec:related}
Prior work on equivalence checking does not support GPU parallelism or synchronization~\cite{counter,erlang,alive2,ddec,cove,stoke,regverif,pnueli,necula,churchill,symdiff,kuncak,ucklee,mirage}.
{\sc Alive}~\cite{alive} symbolically evaluates straight-line single-threaded programs over bit-vectors to generate decision problems for equivalence checking that can be handed over to  SMT solvers. \tool can be seen as generalizing this mechanism to many threads running in parallel that synchronize with each other to perform computations over the reals.
Treating floating-point operations as uninterpreted functions renders standard GPU optimizations unsound~\cite{kleefp,mlirtv}, while fully modeling IEEE-754 semantics is known to cause equivalence checking failures for such optimizations~\cite{realizer}. {\sc Mirage} \citep{mirage} operates at a higher level than {\sc Volta}, directly on a language of multi-linear operators, division, and exponentiation. Ultimately, the final verification conditions look quite similar to \tool's, but with the allowance of rational functions. {\sc Mirage} tackles these conditions via the identity-testing procedure of \citet{li-wu-itcs26}, which, unlike our algorithm, lacks provable soundness (see Section~\ref{sec:decpro}).

Several existing approaches bound the worst-case numerical error in floating-point implementations~\cite{rosa,herbie,lee16,lee18,fptaylor}. However, the worst-case errors when arithmetic operations are reordered in tensor computations (as in GPU optimization) are huge for contrived inputs and do not provide useful insights. For ensuring memory safety in floating-point programs, enforcing non-interference between floating-point and non-floating-point values has proven useful~\cite{gkinder}.

During its operation, the equivalence checker also establishes well-synchronization, guaranteeing both deadlock freedom and the absence of data races. There is a rich body of work that focuses exclusively on data race detection in GPU programs. Traditional data race analyses consider only {\em where} memory accesses occur, ignoring {\em what} data is being accessed, whereas an equivalence checker must reason about both. FaialAA~\cite{faialaa} represents the state of the art in sound data race freedom checking, but, like most prior work, supports only basic synchronization primitives such as {\tt syncthreads} and does not support {\tt syncwarp}. A notable exception is {\sc Weft}~\cite{weft}, which addresses producer-consumer synchronization that lies outside the scope of our work. \tool achieves higher precision than FaialAA because its symbolic evaluation aggressively propagates constant values; for example, it does not report the spurious data race identified by FaialAA in its Alarm 4, Figure 1~\cite{faialaa} (Section~\ref{sec:dr}). However, FaialAA provides sound (though incomplete) analysis for GPU kernels beyond structured-CTAs, including kernels with non-trivial control flow and statically unknown array indices, which \tool rejects outright. The primary advantages of \tool's race checker are its simplicity and efficiency: it runs online during symbolic execution, maintaining only the small additional context $\mathcal{X}$ of Section~\ref{sec:symex}, and adds time linear in the number of executed instructions. Other approaches either suffer from exponential blowup~\cite{gklee,sesa}, fail to detect certain data races~\cite{scord,barracuda,haccrg,ld,curd,dr1,simulee,grace,gmrace}, or exhibit high false positive rates~\cite{vericuda,vercors}. Several analyses rely on reductions~\cite{faialfmsd,faialaa,pug1,pug2,faial,gpuverify} that are sound only when the synchronization primitive is a barrier across all threads. In these methods, the kernel analysis can be partitioned into {\em barrier intervals}, defined as code segments between two consecutive executions of {\tt syncthreads}. However, in the presence of {\tt syncwarp} and tensor core operations, which synchronize only subsets of threads, the barrier interval assumption no longer holds. Finally, {\em test amplification}~\cite{test-amplification} can be seen as a special case of \tool's data race checking, where all threads run identical commands and synchronize exclusively with barriers across all threads.

Lifting systems seek to translate code in low-level, typically general purpose, languages into code in higher-level, typically domain specific, languages so that code can be more easily maintained and better re-optimized, retargeted, or parallelized by compilers. \textsc{C2TACO} \citep{c2taco} lifts C to \textsc{TACO}, as its name suggests, \textsc{STNG} \citep{stng} lifts Fortran to a stencil DSL and then translates to Halide, \textsc{Tenspiler} \citep{tenspiler} lifts C, \CC{}, or Python to a tensor IR and then translates to various backends, such as PyTorch, and \textsc{LLMLift} \citep{llmlift} lifts C, \CC{}, or Python to a custom Python DSL and then translates to a similar set of backends. \textsc{C2TACO} does not provide formal guarantees about its output, instead relying on random testing. However, the latter three systems do verify equivalence of the lifted code and the original code. The fact that these approaches rely on recovering a high-level IR complicates the kinds of VCs they must generate---all of them make non-trivial use of general purpose SMT solvers---and the process of generating the VCs itself, which is done, e.g., in the case of \textsc{STNG}, by counterexample-guided inductive synthesis \citep{cegis}. Moreover, these systems lift sequential code---C, \CC{}, Fortran, or Python loop nests---whereas a GPU kernel comprises hundreds or thousands of concurrent threads. Applying lifting in our setting would additionally require modeling shared memory, barrier and warp-level synchronization, and the attendant possibility of data races and deadlocks.

\section{Conclusion and Future Work}
\label{sec:conclusion}
We have presented \tool, the first equivalence checker for ML GPU kernels. \tool operates directly on low-level PTX, taking advantage of known tensor sizes and thread IDs to simplify the space of expressions it must support. PTX is currently the lowest-level, public-facing language for NVIDIA GPUs and, therefore, the ultimate target of open-source compilation pipelines.

\tool interprets kernels to obtain a symbolic expression for each output tensor element and checks the implied verification conditions using a specialized canonicalization procedure. Soundness and completeness are guaranteed by virtue of the semantics given in this paper and the \textsc{Agda}-verified confluence theorem. We also prove the decidability of identities between certain polynomials with exponentials over the reals.

This paper focuses on verification of programs that synchronize with blocking operations. We leave asynchronous CUDA intrinsics such as {\tt pipeline} and {\tt tma} for future work.

\section*{Data-Availability Statement}

The kernels we used in our benchmarks and the \textsc{Agda} code that proves Lemma~\ref{lem:confluence} are available via Zenodo \cite{artifact-zenodo}, and were submitted for artifact evaluation. Our implementation is available at \url{https://github.com/willtunnels/volta} or as another Zenodo upload \cite{volta-zenodo}.

\begin{acks}

This material is based upon work supported by the Defense Advanced Research
Projects Agency (DARPA) under Contract HR0011-25-2-0039, as part of the MOCHA
program. We would like to thank the following people for their helpful discussions: Jubi
Taneja, Madanlal Musuvathi, Aditya Nori, Ashish Panwar, Aseem Rastogi, Tyler
Sorensen, Nikhil Swamy, Srikar Veluvali, Hari VK, Michael Bauer, Manolis
Papadakis, and Sean Treichler.

\end{acks}

\bibliographystyle{ACM-Reference-Format}
\bibliography{references}

\end{document}